\documentclass[aps,prb,twocolumn]{revtex4}
\usepackage{graphicx,amssymb,amsfonts,amsmath}

\begin{document}

\title{Static and dynamic properties of crystalline phases of two-dimensional electrons
\\
in a strong magnetic field}
\author{A. M. Ettouhami}
   \altaffiliation[Present address: ]{Department of Physics, University of Toronto,  60 St. George St., Toronto M5S 1A7, Ontario, Canada}
   \email{mouneim@physics.utoronto.ca}
\author{F.D. Klironomos}
   \altaffiliation[Present address: ]{Department of Physics, University of California, Riverside, CA 92521}
   \email{fkliron@physics.ucr.edu}
\author{Alan T. Dorsey}
   \email{dorsey@phys.ufl.edu}
\affiliation{Department of Physics, University of Florida, P.O. Box 118440, Gainesville, FL 32611-8440}

\date{\today}


\begin{abstract}

We study the cohesive energy and elastic properties as well as normal modes 
of the Wigner and bubble crystals of the two-dimensional electron system (2DES) in
higher Landau levels. Using a simple Hartree-Fock approach, we show
that the shear moduli ($c_{66}$'s) of these electronic crystals show 
a non-monotonic behavior as a function of the partial filling factor 
$\nu^*$ at any given Landau level, with $c_{66}$ increasing for small 
values of $\nu^*$, before reaching a maximum at some intermediate 
filling factor $\nu^*_m$, and monotonically decreasing for $\nu^*>\nu^*_m$. 
We also go beyond previous treatments, and study how the phase diagram and 
elastic properties of electron solids are changed by the effects of screening by electrons
in lower Landau levels, and by a finite thickness of the experimental sample.
The implications of these results on microwave resonance experiments
are briefly discussed.

\end{abstract}

\maketitle

\section{Introduction}

There has been much interest recently, both theoretically and
experimentally, in the quantum phases of the two-dimensional
electron system in higher Landau levels (LLs). Hartree-Fock (HF)
studies\cite{Fogler1996,Moessner1996} have shown that, for small
partial filling  factors $\nu^* = \nu - 2n\approx 0.1-0.2$ (with
$\nu$ the total filling factor and $n$ the Landau level index), the
electrons form a triangular Wigner crystal (WC), while for $\nu^*$
close to $1/2$ the ground state of the system is a unidirectional
charge density wave or ``stripe'' state.\cite{Fogler1996}
Between these two regions, HF\cite{Fogler1996,Cote2003,Goerbig2004}
and density matrix renormalization group (DMRG)
methods,\cite{Shibata2001} as well as exact diagonalization on small
systems,\cite{Haldane2000} have suggested the existence of a new
kind of crystal structure, the ``bubble crystal'' (BC), with more
than one electron per lattice cell. While HF predicts transitions
between bubble phases with increasing number $M$ of electrons per
bubble, with $M$ going up to $M=n+1$ electrons in LL $n$, in DMRG
the last bubble phase is absent, and a transition takes place
directly from the BC with $M=n$ electrons per bubble to the stripe
state. It is noteworthy that the difference in cohesive energy
between all these different phases remains very small, of order
$0.01-0.03e^2/\epsilon\ell$ (with $e$ the electronic charge,
$\epsilon$ the dielectric constant of the host semiconductor and
$\ell$ the magnetic length).

On the experimental side, evidence for the existence of these
different phases is found mainly through transport experiments.
Early DC measurements\cite{Lilly1999,Du1999} have shown strongly
anisotropic transport around half-filling, which was interpreted as
evidence for a stripe state. Crystal phases of electrons, which are
pinned by quenched disorder, should all be insulating in the DC
regime. One has therefore to resort to other experimental techniques
to resolve the transitions between these different crystal phases
(e.g. WC to BC). One particular such technique consists in measuring
the microwave response of the 2DES, which would give different
resonant behavior for different electronic crystalline phases.
While early microwave conductivity
experiments\cite{Chen2003,Lewis2003} gave considerable support to
the existence of a WC around integer filling $\nu=1,2,3$ and $4$ for
$\nu^*<1/2$, it was not until recently that coexistence between two
phases with two distinct resonant peaks was observed by Lewis {\em
et al.}\cite{Lewis2004}. It is our aim in this paper to discuss in
some detail the experimental results of this last reference, in
light of recent advances in resonance pinning theories of
two-dimensional electronic solids.\cite{Chitra2001,Fogler2000} In
order to do so, we will need to develop a microscopic picture of the
2DES as a system of interacting {\em guiding centers}. Indeed, after
projection of the electronic density on the uppermost Landau level
the problem at hand reduces to one of interacting guiding centers
(of quenched kinetic energy) with an appropriately defined effective
interaction potential that includes quantum effects at the
Hartree-Fock level.
The knowledge of the effective interaction potential between guiding
centers will allow us to find the elastic moduli and the normal
eigenmodes of the Wigner and bubble crystals. In agreement with the
results of Maki and Zotos\cite{Maki1983} for the WC in the lowest
LL, we find that the shear moduli ($c_{66}$'s) of the Wigner and
bubble crystals show a non-monotonic behavior as a function of the
partial filling factor $\nu^*$ at any given Landau level, with
$c_{66}$ increasing for small values of $\nu^*$, before reaching a
maximum at some intermediate filling factor $\nu^*_m$, and then
monotonically decreasing for $\nu^*>\nu^*_m$. These results will
allow us to attempt a qualitative analysis of recent microwave
experiments by Lewis {\em et al.}\cite{Lewis2004}. We find that,
while the $\nu^*$ behavior of the first resonance peak observed in
this last reference is in good qualitative agreement with existing
theoretical predictions for the pinned Wigner
crystal,\cite{Chitra2001} the second resonance peak has a behavior as a function of $\nu^*$
which is quite different from what a linearized solution of
self-consistent equations for the microwave
response theories predict for the two-electron bubble state. 
Below, we shall argue that an adequate description of the
resonance peak of bubble crystals may require the full numerical
solution of the self-consistent response equations, as done
in Ref. \onlinecite{Cote2005}.

This paper is organized as follows. In Sec. \ref{Review} we review
the physics of the Wigner crystal and bubble states, and introduce
some notation. In Sec. \ref{Elasticity} we find the shear moduli of
the Wigner and bubble crystals in LLs $n=2$ and $n=3$. In Sec. \ref{NormalModes}
we calculate the normal modes of the Wigner crystal and of the
two-electron bubble solid in the Landau level $n=2$. 
Then, in Sec. \ref{FiniteThickness} we investigate the effect of 
screening by lower LLs and of finite sample thicknesses on the 
cohesive energies and shear moduli of the Wigner and bubble crystals
in LLs $n=2$ and $n=3$. In Sec.\ref{Consequence} we discuss the implications of our results for the
shear moduli on the microwave response of electronic crystals in
light of the recent microwave conductivity experiments of Lewis {\em
et al.}\cite{Lewis2004} Finally, Sec. \ref{Discussion} contains a
summary of our results along with our conclusions.

\section{Wigner crystal and bubble phases in higher Landau levels: Hartree-Fock approach}
\label{Review}

We shall start from the expression of the Hartree-Fock Hamiltonian of the $n$-th partially filled
Landau level, which is given by (throughout this paper, we use $\int_{\bf q}$ as a shorthand for
$\int \frac{d^2\bf q}{(2\pi)^2}$)
\begin{equation}
H_{HF} = \frac{1}{2} \int_{\bf q} V_{HF}({\bf q})|\rho({\bf q})|^2,
\label{EHF}
\end{equation}
where $V_{HF}$ is the Hartree-Fock interaction, and $\rho({\bf q})$ is the projection of the
electronic density $n({\bf q})$ onto the uppermost LL, and is given by
\begin{equation}
\rho({\bf q}) = \frac{n({\bf q})}{e^{-q^2\ell^2/4}L_n(q^2\ell^2/2)}.
\label{rho(q)}
\end{equation}
For a bubble crystal with $M$ electrons per bubble, we shall approximate the electronic density by
\begin{equation}
n({\bf r}) = \sum_{i}\sum_{m=0}^{M-1} |\varphi_{n,m}({\bf r} - {\bf R}_i)|^2,
\label{electrondensity}
\end{equation}
where $\varphi_{n,m}({\bf r})$ is the noninteracting wavefunction of
angular momentum $m$ and Landau level index $n$, and where the
summation extends over all the bubbles located at the lattice sites
${\bf R}_i$ of a triangular Bravais lattice and over all the
electrons within each bubble. The Hartree-Fock interaction potential
in Eq.~(\ref{EHF}) consists of the sum of a Hartree and Fock parts,
which are given, in the Landau gauge ${\bf A}=Bx\hat{\bf y}$ (${\bf A}$ being the vector potential), by \cite{Cote2003,Goerbig2004}
\begin{subequations}
\begin{gather}
V_H({\bf q})=\frac{2\pi e^2}{\epsilon q}\mbox{e}^{-q^2\ell^2/2}\big[L_n(q^2\ell^2/2)\big]^2,
\label{VH}\\
V_F({\bf q})=- (2\pi\ell^2)\;\int_{{\bf q}'}V_H({\bf q}')e^{-i{\bf q}\times{\bf q}'\ell^2},
\label{VF}
\end{gather}
\end{subequations}
where $L_n(x)$ is the $n$-th Laguerre polynomial.

\begin{figure}[b]
\centering
\includegraphics[angle=-90,totalheight=4.7cm,width=8.09cm,viewport=5 5 555 725,clip]{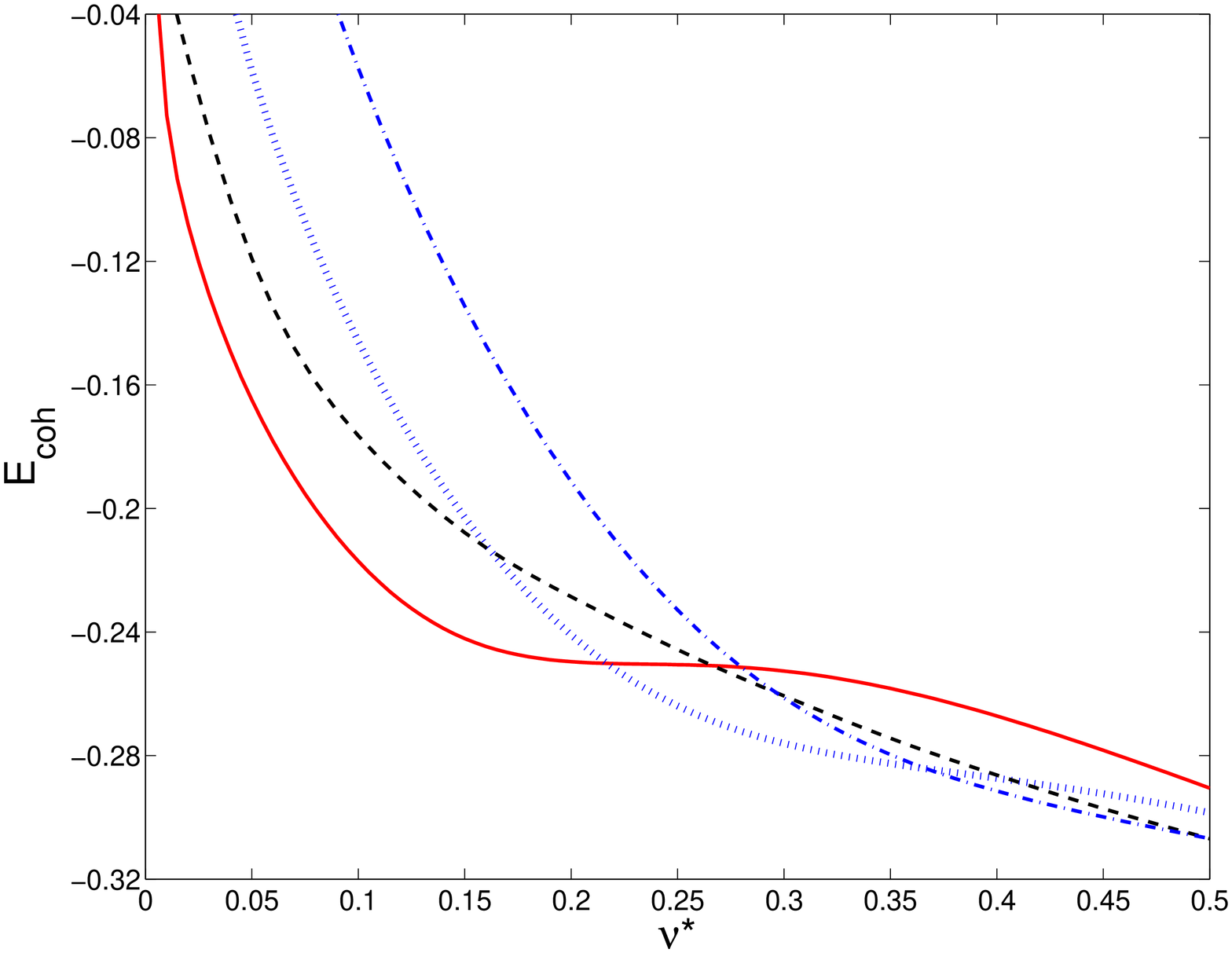}
\includegraphics[angle=-90,totalheight=4.7cm,width=8.09cm,viewport=5 5 555 725,clip]{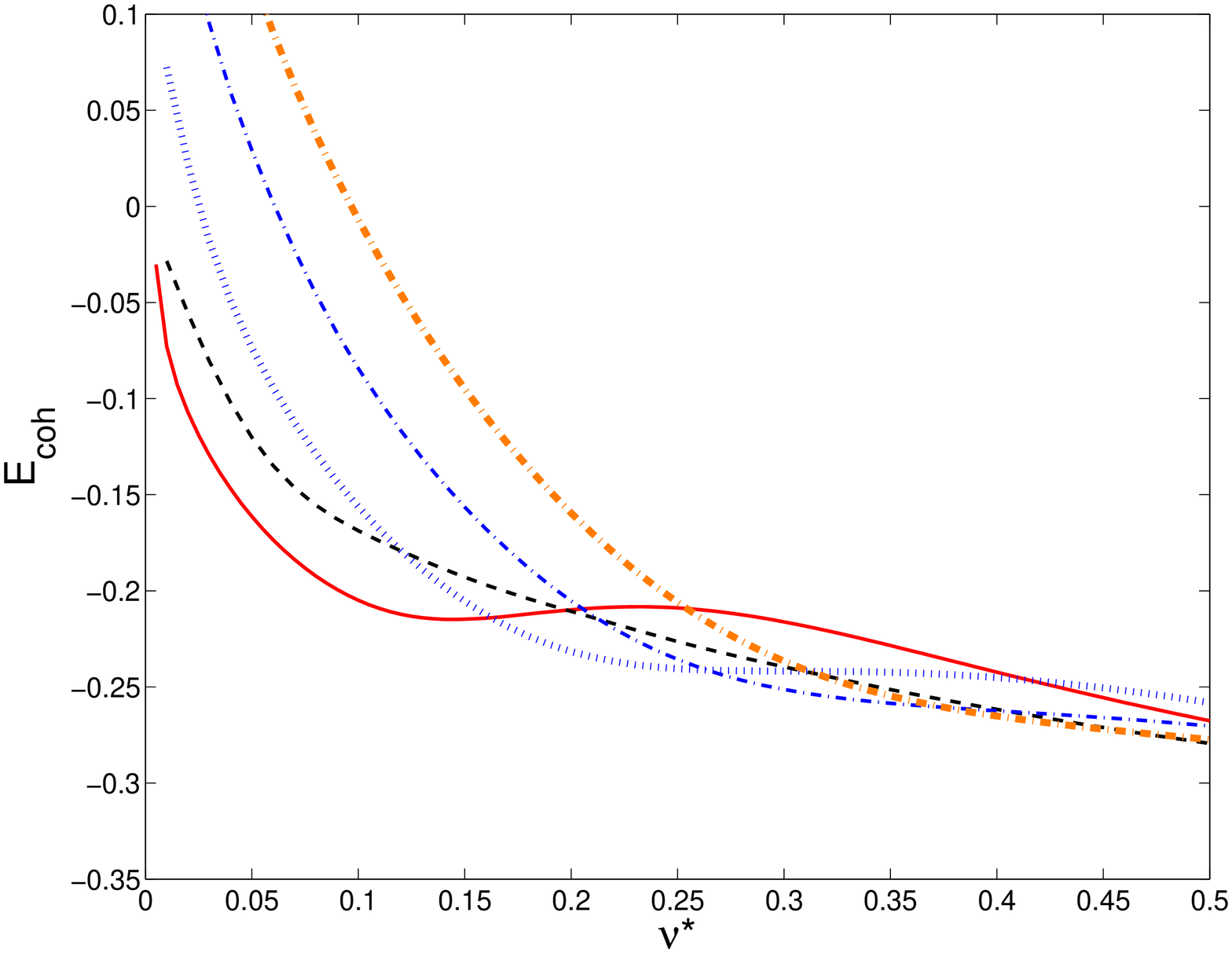}
\caption{(Color online) Cohesive energy of the Wigner crystal (solid line), $M=2$ (dotted line),
$M=3$ (thin dash-dotted line) and $M=4$ bubble solids (thick dash-dotted line in the lower panel).
The cohesive energy of the stripe state is indicated by a dashed line, and all cohesive energies are in units of $e^2/\epsilon\ell$.
The upper panel is for $n=2$, and the lower panel is for $n=3$.}
\label{coh2}
\end{figure}

The calculation of the cohesive energy of the Wigner crystal and bubble phases proceeds in a standard
way as follows. From Eqs. (\ref{rho(q)}) and (\ref{electrondensity}), it is easy to see that we can
write the projected density $\rho({\bf q})$ in the form
\begin{equation}
\rho({\bf q}) = \tilde\rho({\bf q})\sum_i e^{-i{\bf q}\cdot{\bf R}_i},
\label{rhotilderho}
\end{equation}
where the sum runs through all vectors ${\bf R}_i$ of the triangular Bravais lattice, and where
$\tilde\rho({\bf q})$ is given by
\begin{equation}
\tilde\rho({\bf q}) = \frac{\tilde{n}({\bf q})}{e^{-q^2\ell^2/4}L_n(q^2\ell^2/2)},
\label{tilderho}
\end{equation}
with $\tilde{n}({\bf q})$ the Fourier transform of the density at a given bubble,
\begin{equation}
\tilde{n}({\bf q})=\sum_{m=0}^{M-1}\int d{\bf r}|\varphi_{nm}({\bf r})|^2e^{-i{\bf q}\cdot{\bf r}}.
\label{tilden}
\end{equation}
Inserting the decomposition (\ref{rhotilderho}) into the expression of the cohesive energy,
Eq.~(\ref{EHF}), and making use of Poisson's summation formula (here $A_c= 2\pi\ell^2 M/\nu^*$ is the area of the unit
lattice cell and the ${\bf Q}$'s are reciprocal lattice vectors)
\begin{equation}
\sum_i e^{-i{\bf q}\cdot{\bf R}_i}=\frac{(2\pi)^2}{A_c}\sum_{\bf Q}\delta({\bf q}-{\bf Q}),
\label{Poisson}
\end{equation}
we finally obtain the following expression for the cohesive energy per particle
$E_{coh}=\langle H_{HF}\rangle/N$~:
\begin{equation}
E_{coh}=\frac{1}{2A_c}\sum_{\bf Q}\big[V_H({\bf Q})(1-\delta_{{\bf Q},{\bf 0}}) + V_F({\bf Q})
\big]|\tilde{\rho}({\bf Q})|^2,
\end{equation}
where charge neutrality requires that we exclude the diverging ${\bf Q}=0$ 
Hartree term from the summation. In Fig.~\ref{coh2}, we show
the cohesive energies of the different phases (Wigner crystal,
bubble solids and stripes) in the Landau levels $n=2$ and $n=3$. The
values of the cohesive energies shown, and of the partial filling
factor $\nu^*$, at which transitions between the different phases
occur in both Landau levels, are in excellent agreement with recent
Hartree-Fock calculations by C\^ot\'e {\em et al.}\cite{Cote2003}
and by Goerbig {\em et al.}\cite{Goerbig2004}
As it can be seen, the energy difference between the WC, BC and stripe
states is extremely small, of order $0.02e^2/\epsilon\ell$, which is within the range of
the magnetophonon and magnetoplasmon excitation spectrum of these phases. 
It is therefore conceivable that quantum
fluctuations, in the form of zero-point energy, may alter the cohesive energies and
therefore change the structure of the phase diagram of the 2DES in higher LLs.
This point will be investigated in Sec. \ref{NormalModes}, 
where we shall write an equation of motion for the electron guiding
centers in the Wigner and bubble crystals, with the
goal of finding the normal eigenmodes (and hence the zero-point energy) of these structures.
An essential ingredient in such a calculation is the elastic matrix of the Wigner
and bubble crystals, that we shall determine in Sec. \ref{Elasticity} below.

\section{Elastic matrix and elastic moduli of the Wigner and bubble crystals in higher Landau levels}
\label{Elasticity}

Having reviewed the basic phase diagram of the 2D electron gas in
partially filled LLs, we now would like to study the elastic
properties of the Wigner and bubble crystals. But, in order to be
able to do so, we still need to derive an effective interaction
between the guiding centers of the electrons, which in turn will
allow us to derive the elastic matrix of the WC and BC, and hence
find the compression and shear moduli of these crystalline
structures. This will be the object of the following Subsection.

\subsection{Effective interaction potential between guiding centers}
\label{EffInt}

Going back to Eqs.~(\ref{rhotilderho})-(\ref{tilden}), if we furthermore write for the projected density
$\tilde\rho({\bf q})$ the decomposition $\tilde\rho({\bf q})=\sum_m\tilde\rho_m({\bf q})$, with
\begin{equation}
\tilde\rho_m({\bf q}) = \frac{\int d{\bf r}|\varphi_{nm}({\bf r})|^2e^{-i{\bf q}\cdot{\bf r}}}
{e^{-q^2\ell^2/4}L_n(q^2\ell^2/2)},
\end{equation}
then we can rewrite the cohesive energy $E_{HF}$ in the form
\begin{equation}
E_{HF}=\frac{1}{2}\sum_{\underset{m,m'}{i\neq j}}U_{mm'}({\bf R}_i-{\bf R}_j)+\!\!\!\sum_{i,m<m'}U_{mm'}(0),
\end{equation}
where we introduced the effective interaction potential $U_{mm'}$ between {\em guiding centers} of electrons in states
$m$ and $m'$, which in real space is given by
\begin{equation}
U_{mm'}({\bf r})=\int_{\bf q}\tilde{\rho}_m({\bf q})V_{HF}({\bf q})\tilde{\rho}_{m'}({\bf q})
e^{i{\bf q}\cdot{\bf r}}.
\label{Ueff}
\end{equation}
In the following Subsection, we shall use the above expression of the interaction potential to study the
normal modes of the bubble crystal in higher ($n\ge 2$) LLs.

\subsection{Elastic moduli of the electron crystal}
\label{ElasticModuli}

The derivation of the elastic moduli associated with the effective
interaction potential in Eq. (\ref{Ueff}) proceeds in a standard way
as follows. First, we evaluate the elasticity matrix
$\Phi_{\alpha\beta}({\bf R})$, which is given by\cite{Ashcroft1988}
\begin{equation}
\Phi_{\alpha\beta}({\bf R})=\delta_{{\bf R},{\bf 0}}\sum_{k}\partial_{\alpha}\partial_{\beta}U({\bf R}_k)
-\partial_{\alpha}\partial_{\beta}U({\bf R}),
\end{equation}
with $U$ the total interaction potential between bubbles:
\begin{equation}
U({\bf r})=\sum_{m,m'} U_{mm'}({\bf r}).
\end{equation}

Using the following definition of the direct and
inverse Fourier transformations (here the ${\bf q}$ integration is carried over the first Brillouin zone
of the reciprocal lattice, and we remind the reader that $A_c$ is the area of the primitive unit cell of
the bubble crystal):
\begin{subequations}
\begin{eqnarray}
\Phi_{\alpha\beta}({\bf q}) & = & \sum_{\bf R}
\Phi_{\alpha\beta}({\bf R})e^{-i{\bf q}\cdot{\bf R}}, \\
\Phi_{\alpha\beta}({\bf R}) & = & \frac{1}{A_c}\int_{\bf q} \Phi_{\alpha\beta}({\bf q})
e^{i{\bf q}\cdot{\bf R}},
\end{eqnarray}
\end{subequations}
and the identity (\ref{Poisson}), one can easily show that $\Phi_{\alpha\beta}({\bf q})$ can be written
in the form
\begin{eqnarray}
\Phi_{\alpha\beta}({\bf q}) &=&\frac{1}{A_c}\sum_{\bf Q}\big[(q_\alpha +Q_\alpha)(q_\beta + Q_\beta)
U({\bf q}+{\bf Q})
\nonumber\\
&-&Q_\alpha Q_\beta U({\bf Q})\big].
\end{eqnarray}
Expanding the second term on the right hand side of the above equation in $q$ around $q=0$ leads to the
result
\begin{widetext}
\begin{eqnarray}
\Phi_{\alpha\beta}({\bf q})&=&\frac{1}{A_c}q_\alpha q_\beta
\Bigg\{U(q)+\sum_{{\bf Q}\neq 0}
\Bigg[U(Q)+(Q_\alpha^2+Q_\beta^2)\frac{U'(Q)}{Q}+Q_\alpha^2
Q_\beta^2\Big(\frac{QU''(Q)-U'(Q)}{Q^3}
\Big)\Bigg]\Bigg\} \notag\\
&+&\frac{1}{2A_c}\delta_{\alpha\beta}\sum_{{\bf Q}\neq
0}\Big[q^2Q_\alpha^2\frac{U'(Q)}{Q}+Q_\alpha^2
(q_x^2Q_x^2+q_y^2Q_y^2)\Big(\frac{QU''(Q)-U'(Q)}{Q^3}\Big)\Big]
\end{eqnarray}
\end{widetext}
Given that the elastic matrix of a two-dimensional triangular lattice is of the general form
\begin{equation}
\Phi_{\alpha\beta}({\bf q})=(c_{11}-c_{66})q_{\alpha}q_{\beta}
+ c_{66}q^2\delta_{\alpha\beta},
\end{equation}
we see that the compression and shear moduli can be extracted from the
expression of $\Phi_{\alpha\beta}({\bf q})$ according to:
\begin{subequations}
\begin{eqnarray}
c_{11}(q)&=&\frac{\Phi_{xx}(q_x\hat{\bf x})}{q_x^2},\\
c_{66}(q)&=&\frac{\Phi_{xx}(q_y\hat{\bf y})}{q_y^2}.
\end{eqnarray}
\end{subequations}
Using the above definitions, and the fact that:
\begin{equation}
A_c = 2\pi\ell^2 \big(\frac{M}{\nu^*}\big),
\end{equation}
we obtain (note that these moduli have units of energy; in order to obtain
elastic moduli in units of energy per unit area, one has to divide
by the unit cell area $A_c$):
\begin{widetext}
\begin{subequations}
\begin{eqnarray}
c_{11}&=&\frac{\nu^*}{(2\pi M\ell^2)}\Bigg\{ U(q) + \sum_{{\bf
Q}\neq 0} \Bigg[U(Q)+\frac{5}{2}Q_x^2\frac{U'(Q)}{Q}
+\frac{1}{2}Q_x^4 \Big(\frac{QU''(Q)-U'(Q)}{Q^3}\Big)
\Bigg]\Bigg\},\label{c11}
\\
c_{66}&=&\frac{\nu^*}{2(2\pi M\ell^2)}\sum_{{\bf Q}\neq
0}\Bigg[Q_x^2\frac{U'(Q)}{Q} +Q_x^2Q_y^2
\Bigg(\frac{QU''(Q)-U'(Q)}{Q^3}\Bigg)\Bigg]. \label{c66}
\end{eqnarray}
\end{subequations}
\end{widetext}
In the above expressions, the reciprocal lattice vectors ${\bf Q}$ for a triangular lattice are given by
\begin{equation}
{\bf Q}=\frac{2\pi}{a}\Big(\frac{2n-m}{\sqrt{3}}\hat{\bf x}+m\hat{\bf y}\Big),
\label{Eq:Qs}
\end{equation}
where the lattice spacing $a$ is given by
\begin{equation}
a = \Big(\frac{4\pi M}{\sqrt{3}\nu^*}\Big)^{1/2} \ell.
\label{latticecst}
\end{equation}
From Eq.~(\ref{c11}), we see that the long wavelength limit ($q\to 0$) of the compression modulus of the
bubble crystal is given by
\begin{equation}
c_{11}(q\to 0)\sim \frac{\nu^*}{M}\Big(\frac{e^2}{\epsilon\ell}\Big)\frac{1}{q\ell},
\end{equation}
and shows the characteristic $1/q$ plasmon behavior, in agreement
with the well-known result for the two-dimensional classical Wigner
crystal in zero magnetic field.\cite{Bonsall1977}

\begin{figure}[t]
\centering
\includegraphics[angle=-90,totalheight=4.7cm,width=8.09cm,viewport=5 10 555 730,clip]{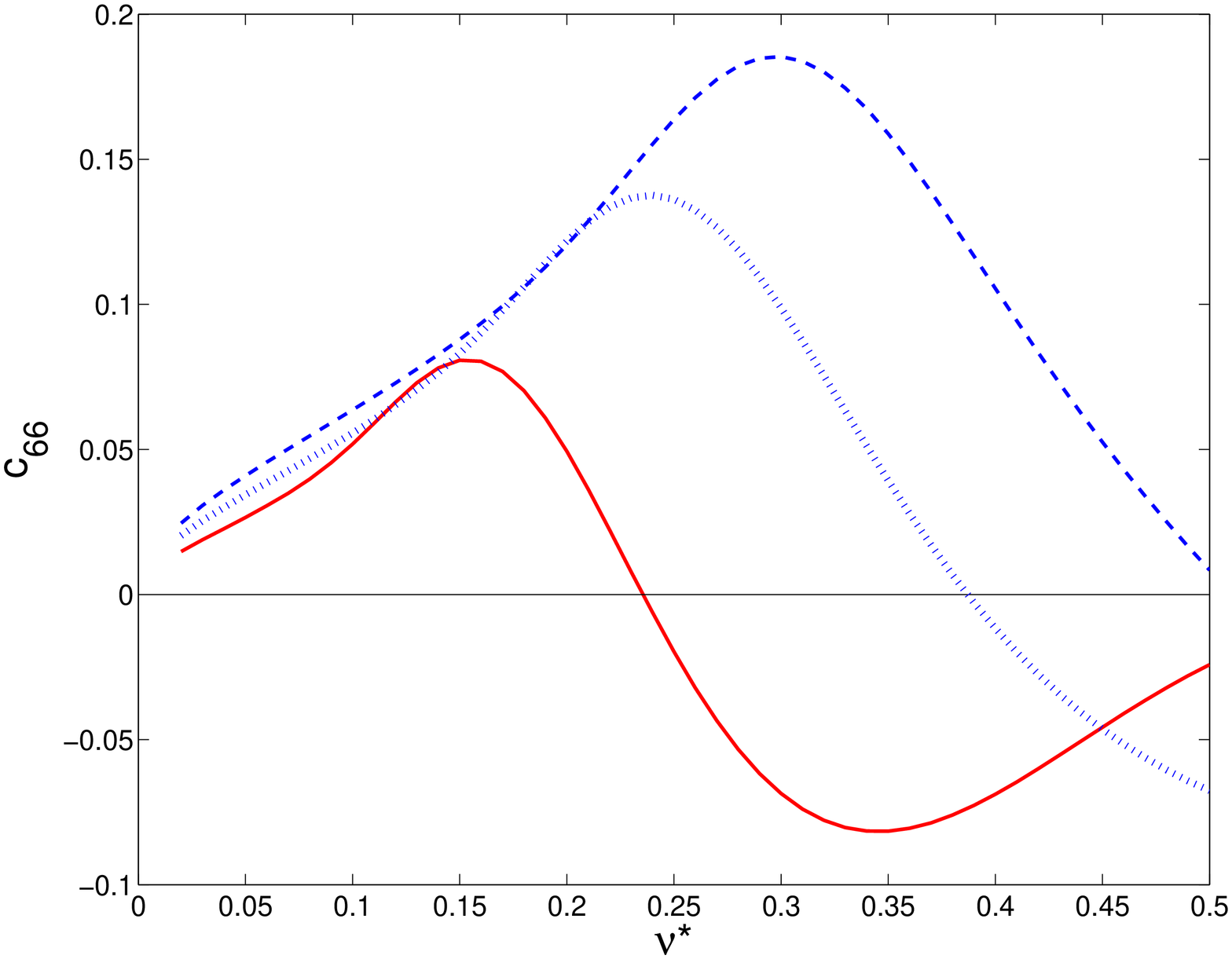}
\includegraphics[angle=-90,totalheight=4.7cm,width=8.09cm,viewport=5 5 560 725,clip]{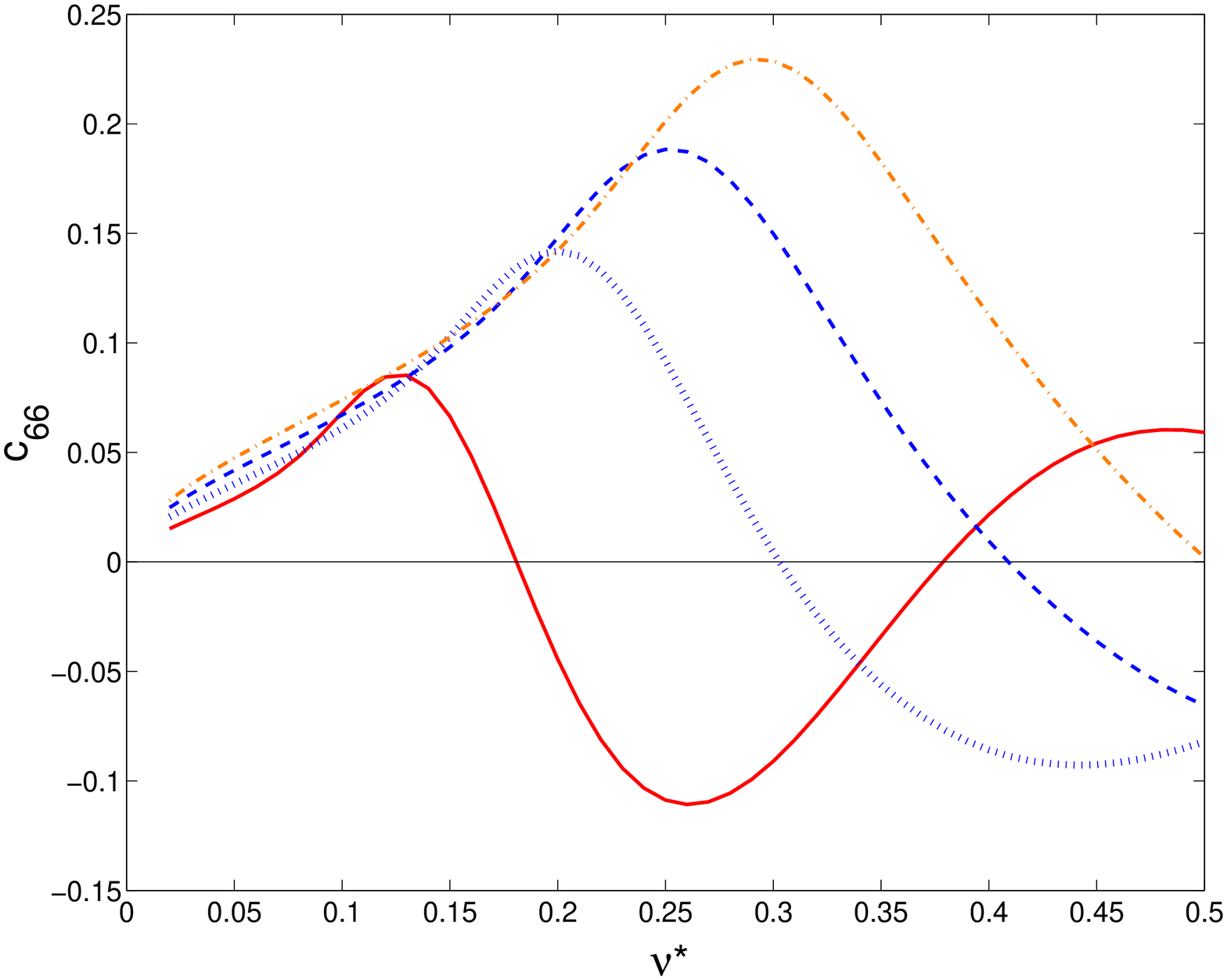}
\caption{(Color online) Upper panel: shear moduli of the various electronic crystals (in units of $e^2/\epsilon\ell$) vs.
$\nu^*$ in the $n=2$ Landau level. The solid line is for the Wigner crystal, the dotted line
is for the 2e BC, and the dashed line is for the 3e BC. 
Lower panel: same thing for the n=3 level. The dash-dotted line is for the $M=4$ bubble solid.}
\label{figc66}
\end{figure}

Figure~\ref{figc66} shows the variation of the shear moduli of
various electronic crystals vs. $\nu^*$ in the Landau levels $n=2$
and $n=3$. For small values of the partial filling factor $\nu^*$,
$c_{66}$ is an increasing function of $\nu^*$. $c_{66}$ reaches a
broad maximum at some characteristic value $\nu^*_m$, and shows a
decreasing behavior for $\nu^*>\nu_m^*$. At some higher value of the
partial filling factor $\nu^*$, the shear modulus turns negative,
indicating an instability of the electronic crystal. Similar results
have been obtained by C\^ot\'e {\em et al.}\cite{Cote2005},
using a rather sophisticated generalized random phase approximation (GRPA) method.
Although the precise location of the maxima of the shear modulus curves
is slightly different in their case, the general qualitative behavior
of the shear moduli that we find is in good qualitative agreement with
the results of Ref. \onlinecite{Cote2005}. It is in fact quite remarkable
that we were able to reproduce the qualitative behavior of the 
more involved GRPA method using a simple wavefunction ansatz. This
indicates that using single-particle non interacting wavefunctions
is a good starting point approximation to study the physics of 
electronic lattices in quantum Hall systems in higher Landau levels.

\begin{figure}[b]
\centering
\includegraphics[angle=-90,totalheight=4.7cm,width=8.09cm,viewport=5 15 560 795,clip]{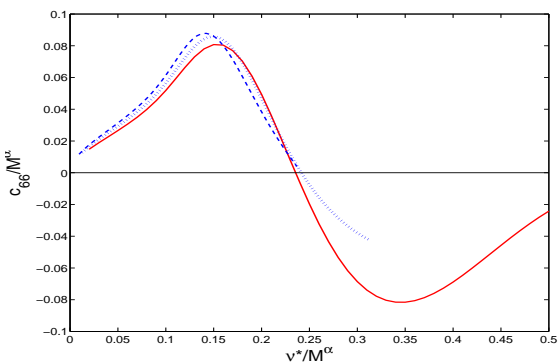}
\caption[]{
Universal scaling of the shear modulus for the Wigner and bubble crystals in the $n=2$ Landau level.
Here the exponent $\alpha\simeq 0.68$.
}
\label{Fig:scaling}
\end{figure}

The peculiar shape of the shear moduli curves in Fig. \ref{figc66} 
suggests that there may be a universal scaling law for
the shear modulus, of the form: 
\begin{equation}
c_{66}(\nu^*) = M^{\alpha} f(\nu^*/M^{\beta}), 
\label{Eq:scaling}
\end{equation}
with a universal function $f$ and some scaling exponents $\alpha$ and $\beta$. 
The identical way in which $c_{66}$ increases at small $\nu^*$ for different values of $M$ 
indicates that $\alpha\simeq \beta$. Figure \ref{Fig:scaling}
shows that, at least for the first peak, this is indeed the case, with exponents
$\alpha = \beta \simeq 0.68$ for all three values of $M$, $M=1,\,2,$ and $3$.

In Sec. \ref{Consequence}, we will use the results of this Section as an input to
the variational replica theory of Chitra et {\em al.}\cite{Chitra2001} 
and try to analyze the experimental data of Lewis {\em et al.}\cite{Lewis2004} 
for the microwave conductivity of the two dimensional electron system 
in the $n=2$ Landau level. For the moment, we shall turn our attention to the normal
modes of the Wigner and bubble crystals in higher Landau levels.

\section{Normal modes and zero point energy of the Wigner and bubble crystals}
\label{NormalModes}

\begin{figure}[t]
\centering
\includegraphics[angle=-90,totalheight=4.7cm,width=8.09cm,viewport=5 15 560 795,clip]{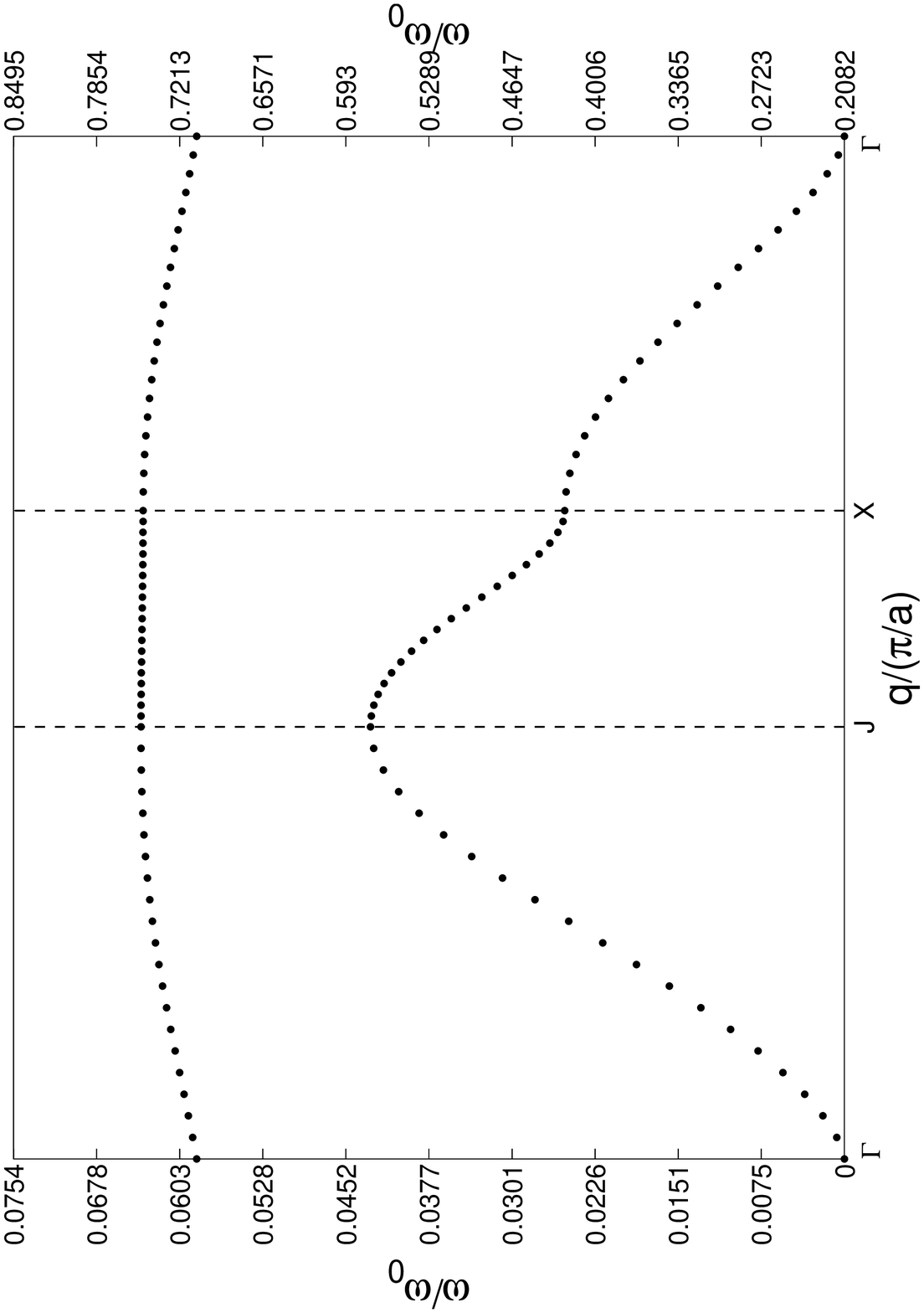}
\includegraphics[angle=-90,totalheight=4.7cm,width=8.09cm,viewport=5 15 560 770,clip]{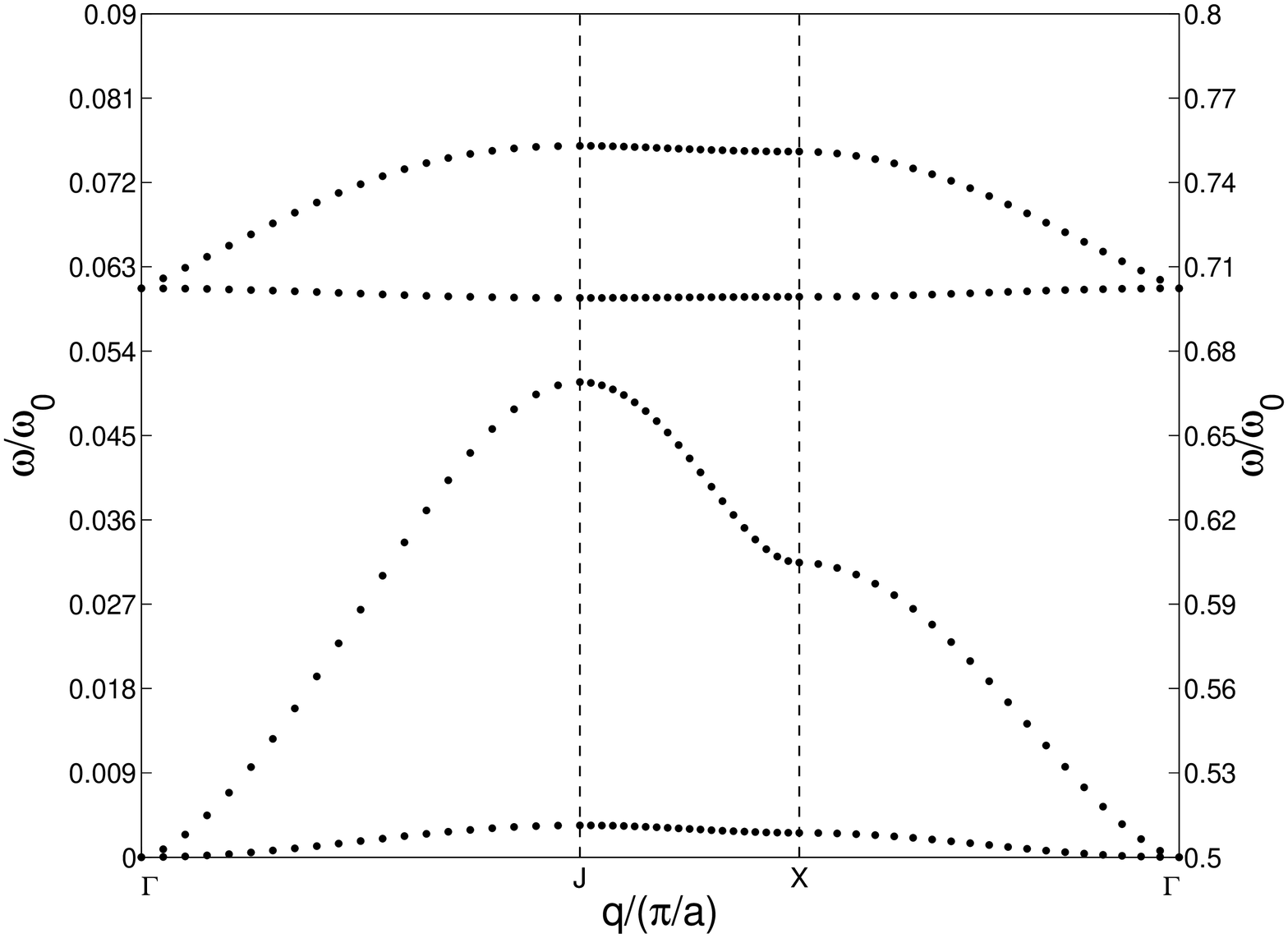}
\caption[]{
Magnetophonon and magnetoplasmon dispersion curves along the boundary $\Gamma-J-X-\Gamma$ of the
irreducible element of the first Brillouin zone for the WC at $\nu^*=0.18$ (upper panel) and
of the $M=2$ BC at $\nu^*=0.25$ (lower panel). Both plots are for
the $n=2$ Landau level. Left axes correspond to magnetophonon values and right axes to magnetoplasmon
values, and are all measured in units of $\omega_0=e^2/\epsilon\hbar\ell$.}
\label{NM-WCBC}
\end{figure}

We now turn our attention to the derivation of the normal modes of the Wigner and bubble crystals. 
To fix ideas, we shall consider the simplest case of the two-electron bubble crystal in the $n=2$ Landau
level, and write for the electron guiding centers ${\bf r}_{mi}$ ($m=0,1$  being the angular momentum
quantum number) at a given lattice site ${\bf R}_i$ the decomposition
${\bf r}_{mi}={\bf R}_i+{\bf u}_{mi}$, where ${\bf u}_{mi}$ is the displacement of the $m$-th electron
from the equilibrium lattice position ${\bf R}_{i}$. Expanding the HF energy of a distorted Wigner
or bubble crystal
\begin{eqnarray}
E_{HF}&=&\frac{1}{2}\sum_{\underset{m,m'}{i\neq j}}U_{mm'}({\bf r}_{mi}-{\bf r}_{m'j})
\nonumber\\
&+&\sum_{i,m<m'}U_{mm'}({\bf u}_{mi}-{\bf u}_{m'i})
\end{eqnarray}
to second order in the small displacements ${\bf u}_{mi}$ leads to the elastic energy
\begin{equation}
E_{el} = \frac{1}{2}\sum_{\underset{m,m'}{i,j}}u_{m\alpha}({\bf R}_i)\tilde\Phi^{mm'}_{\alpha\beta}
({\bf R}_i-{\bf R}_j)u_{m'\beta}({\bf R}_j),
\end{equation}
where the generalized elastic matrix $\tilde\Phi$ has the following expression
\begin{eqnarray}
\tilde\Phi^{mm'}_{\alpha\beta}({\bf R}_i-{\bf R}_j)&=&\delta_{ij}\delta_{mm'}\sum_{k,m''}\partial_{\alpha}
\partial_{\beta}U_{mm''}({\bf R}_k)\notag\\
&-&\partial_{\alpha}\partial_{\beta}U_{mm'}({\bf R}_i-{\bf R}_j).
\label{elmat}
\end{eqnarray}
Note that, by contrast to the previous Section, where all the electrons within the same
bubble were assumed to have the same displacement vector ${\bf u}_i$, here we
consider the more general case of electrons within the bubble moving independently from one another.
This leads to an elastic matrix $\tilde\Phi_{\alpha\beta}^{mm'}$ which depends on the internal quantum numbers $m,\, m'$ of electrons
within bubbles. This last elastic matrix is related to the elastic matrix $\Phi_{\alpha\beta}$ of the previous section through
the equation:
\begin{equation}
\Phi_{\alpha\beta} = \sum_{m,m'}\tilde\Phi_{\alpha\beta}^{mm'}.
\end{equation}
We now are in a position to find the normal modes of the bubble
crystal. In the presence of a magnetic field, the equation of motion for the $m$-th electron at
lattice site ${\bf R}_i$ can be written in the form ($m^*$ is the effective mass of the electron in
the host semiconductor and $\varepsilon_{\alpha\beta}$ is the totally antisymmetric tensor in two
dimensions)
\begin{eqnarray}
m^*\frac{d^2}{dt^2}u_{m\alpha}({\bf R}_i)&=&-\sum_{j,m'}\tilde{\Phi}_{\alpha\beta}^{mm'}({\bf R}_i-{\bf R}_j)
u_{m'\beta}({\bf R}_j)\notag\\
&-&\frac{eB}{c}\varepsilon_{\alpha\beta}\frac{d}{dt}u_{m\beta}({\bf R}_i).
\label{eqmot}
\end{eqnarray}
We shall seek a solution to the above equation of motion that
represents a wave with angular frequency $\omega$ and wavevector
${\bf q}$, {\em i.e.} $u_{m\alpha}({\bf R}_i)=A_{m\alpha}({\bf
q})e^{i({\bf q}\cdot{\bf R}_i-\omega t)},$ where the ${\bf A}_{m}$'s
are complex coefficients whose ratios specify the relative
amplitude and phase of the vibrations of the electrons within each
primitive cell. Substituting the above expression into
Eq.~(\ref{eqmot}) results in the following secular equation for the
Wigner crystal:
\begin{equation}
\begin{pmatrix}
\tilde{\Phi}_{xx}^{00}({\bf q})-\omega^2 & \tilde{\Phi}_{xy}^{00}({\bf q})-i\omega\omega_c \\
\tilde{\Phi}_{xy}^{00}({\bf q})+i\omega\omega_c  & \tilde{\Phi}_{yy}^{00}({\bf q})-\omega^2
\end{pmatrix}
\begin{pmatrix}
A_{0x} \\ A_{0y}
\end{pmatrix} =0,
\label{secEqWC}
\end{equation}
while for the $M=2$ bubble solid the secular equation is given by
\begin{widetext}
\begin{equation}
\begin{pmatrix}
\tilde{\Phi}_{xx}^{00}({\bf q})-\omega^2 & \tilde{\Phi}_{xx}^{01}({\bf q}) &
\tilde{\Phi}_{xy}^{00}({\bf q})-i\omega\omega_c & \tilde{\Phi}_{xy}^{01}({\bf q}) \\
\tilde{\Phi}_{xx}^{01}({\bf q}) & \tilde{\Phi}_{xx}^{11}({\bf q})-\omega^2 &
\tilde{\Phi}_{xy}^{01}({\bf q}) & \tilde{\Phi}_{xy}^{11}({\bf q})-i\omega\omega_c \\
\tilde{\Phi}_{xy}^{00}({\bf q})+i\omega\omega_c & \tilde{\Phi}_{xy}^{01}({\bf q}) &
\tilde{\Phi}_{yy}^{00}({\bf q})-\omega^2 & \tilde{\Phi}_{yy}^{01}({\bf q}) \\
\tilde{\Phi}_{xy}^{01}({\bf q}) & \tilde{\Phi}_{xy}^{11}({\bf q})+i\omega\omega_c &
\tilde{\Phi}_{yy}^{01}({\bf q}) & \tilde{\Phi}_{yy}^{11}({\bf q})-\omega^2
\end{pmatrix}
\begin{pmatrix}
A_{0x} \\ A_{1x} \\ A_{0y} \\ A_{1y}
\end{pmatrix}=0\,,
\label{secEq}
\end{equation}
\end{widetext}
where we defined
\begin{equation}
\tilde{\Phi}_{\alpha\beta}^{mm'}({\bf q})=\frac{1}{m^*}\sum_{i}\tilde{\Phi}_{\alpha\beta}^{mm'}({\bf r})
e^{-i{\bf q}\cdot{\bf R}_i},
\label{dynmat}
\end{equation}
and used the fact that
$\tilde{\Phi}_{\alpha\beta}^{mm'}({\bf q})=\tilde{\Phi}_{\beta\alpha}^{mm'}({\bf q})
=\tilde{\Phi}_{\alpha\beta}^{m'm}({\bf q})$.

The above secular equations have a non-vanishing solution (for the
${\bf A}_{m\alpha}$'s) only if the determinant of the secular matrix
is zero. This leads to two types of solutions for the eigenmodes:\cite{Cote1990}
magnetophonon modes, with eigenfrequencies which vanish like
$q^{3/2}$ as $q\to 0$, and magnetoplasmon modes which tend to a
finite limit as $q\to 0$. In Fig.~\ref{NM-WCBC} we show our solution
for the magnetophonon and magnetoplasmon modes of the WC and the
$M=2$ BC in the $n=2$ Landau level. As is expected, the eigenmode
spectrum for the $2e$ BC has four branches (compared to two for the
WC). We note that the magnetophonon dispersion curves we obtain are
identical to the ones obtained by C\^ot\'e {\em et
al.}\cite{Cote2003} within the time-dependent HF approach. 
To our knowledge, however, the magnetoplasmon modes of electronic crystals, shown here
in Fig. \ref{NM-WCBC}, have not been previously studied in higher LLs.

We now turn our attention to the calculation of the zero point energy of
the Wigner and bubble crystals. To this end, we shall use the
approach of Cunningham,\cite{Cunningham1974} whereby one evaluates
the energy of the magnetophonon and magnetoplasmon modes at a given
number $N_p$ of predefined points ${\bf q}_i$ within the irreducible
element of the first Brillouin zone for the triangular lattice, with
appropriate weights $\alpha_i$ assigned to each one of these points.
The zero-point energy (per electron) for the $M$-electron BC, in units of
$e^2/\epsilon\ell$, is then given by
\begin{equation}
E_{ZP}=\frac{1}{2M\omega_0}\sum_{j=1}^{2M}\sum_{i=1}^{N_p}\alpha_i\omega_j({\bf q}_i),
\end{equation}
where $\omega_0=e^2/\epsilon\hbar\ell$.
The resulting zero-point energies for $N_p=6$ (using the special points and
associated weights\cite{Remark1} given in Ref.
\onlinecite{Cunningham1974}) are shown for a range of partial
filling factor values in Table.~\ref{ZP-energy}.
We notice that for $\nu^*\leq 0.20$ the zero-point energy of the WC always exceeds the corresponding quantity for the
two-electron BC, which suggests that quantum fluctuations will shift the transitional filling factor $\nu^*_{1,2}$ between
the WC and the 2e BC toward smaller values. However,
this shift in the transitional values of $\nu^*$, of order 0.02,
is rather small, and has no substantial effect on the overall phase diagram of the 2DES in higher LLs.

\begin{table}[t]
\begin{tabular}{cccc}
\hline\hline
$\nu^*$ &  $E_{WC}$ &  $E_{2eBC}$ \\ \hline
0.05    &  0.362825 &  0.360756   \\
0.10    &  0.368391 &  0.360716   \\
0.15    &  0.379527 &  0.362053   \\
0.20    &  0.382453 &  0.365264   \\
0.25    &  0.367643 &  0.368439   \\ \hline\hline
\end{tabular}
\caption{Zero point energy (per particle) of the WC and 2e BC for different
partial filling factor values $\nu^*$.  Energy is measured
in units of $e^2/\epsilon\ell$. \label{ZP-energy}}
\end{table}

\section{Effect of finite sample thickness and of lower Landau levels}
\label{FiniteThickness}

Let us now investigate the effect of finite sample thickness and of
lower Landau levels on the energetics of the Wigner and bubble crystals in higher LLs. To this end, we
shall replace the bare Coulomb potential in Eqs.~(\ref{VH})-(\ref{VF}) with the
following effective interaction
\begin{equation}
v(q)=\frac{2\pi e^2}{\varepsilon(q)q}e^{-\lambda q\ell}.
\label{newV}
\end{equation}
The parameter $\lambda$ models a finite thickness sample, and is
generally taken to be of order unity.\cite{Zhang1986} The effect of
lower LLs is encoded in the wavevector-dependent dielectric constant
$\varepsilon(q)$, for which we shall use the following expression,
due to Aleiner and Glazman\cite{Aleiner1995}
\begin{equation}
\varepsilon(q) =\epsilon\Big\{1 + \frac{2}{q a_B}[1-J_0^2(qR_c)]\Big\}, \label{dielectricCst}
\end{equation}
with $a_B=\hbar^2\epsilon/m^*e^2$ the effective Bohr radius and $R_c=\sqrt{2n+1}\ell$.

\begin{figure}[t]
\centering
\includegraphics[angle=-90,totalheight=4.7cm,width=8.09cm,viewport=5 5 555 725,clip]{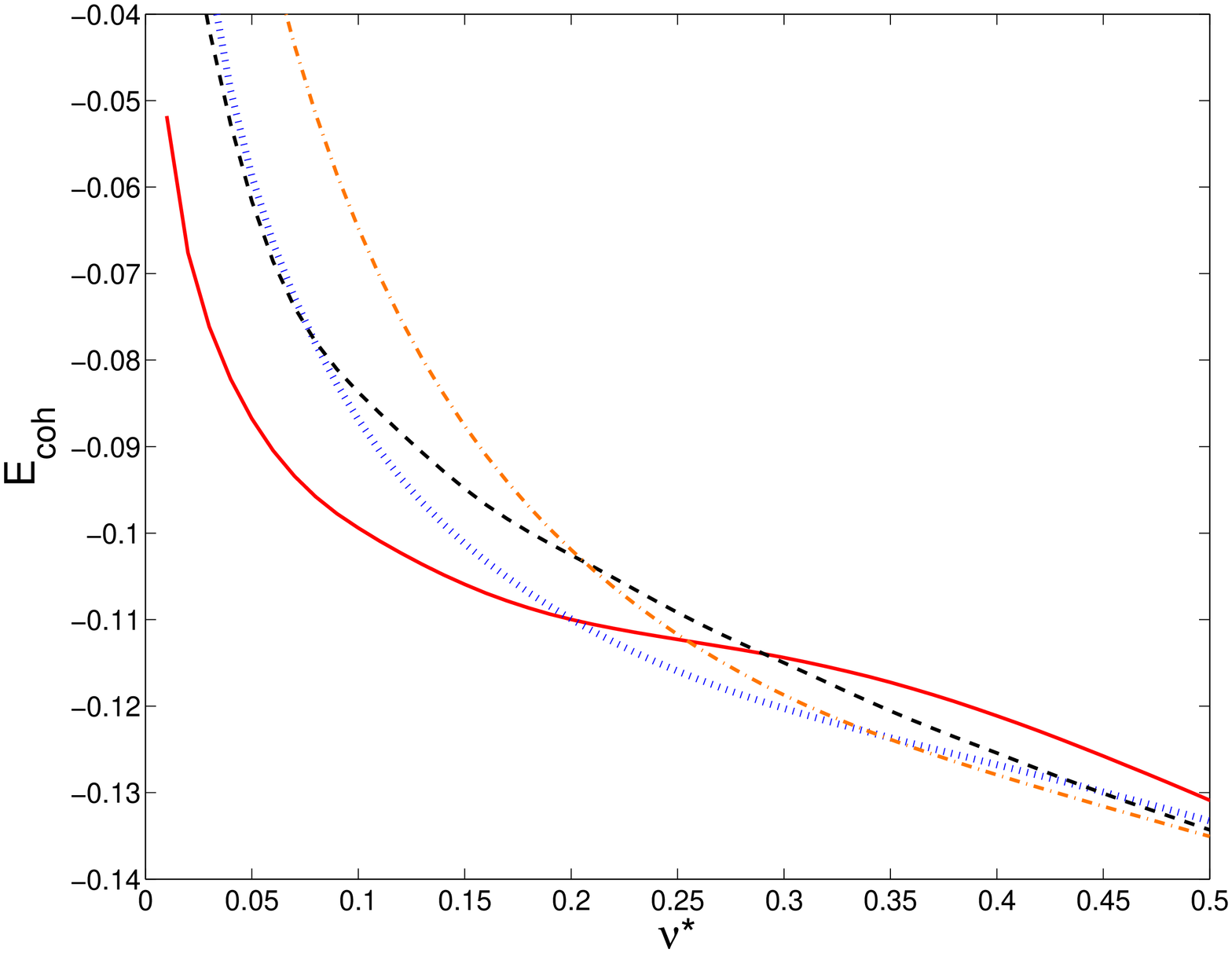}
\includegraphics[angle=-90,totalheight=4.7cm,width=8.09cm,viewport=5 5 555 725,clip]{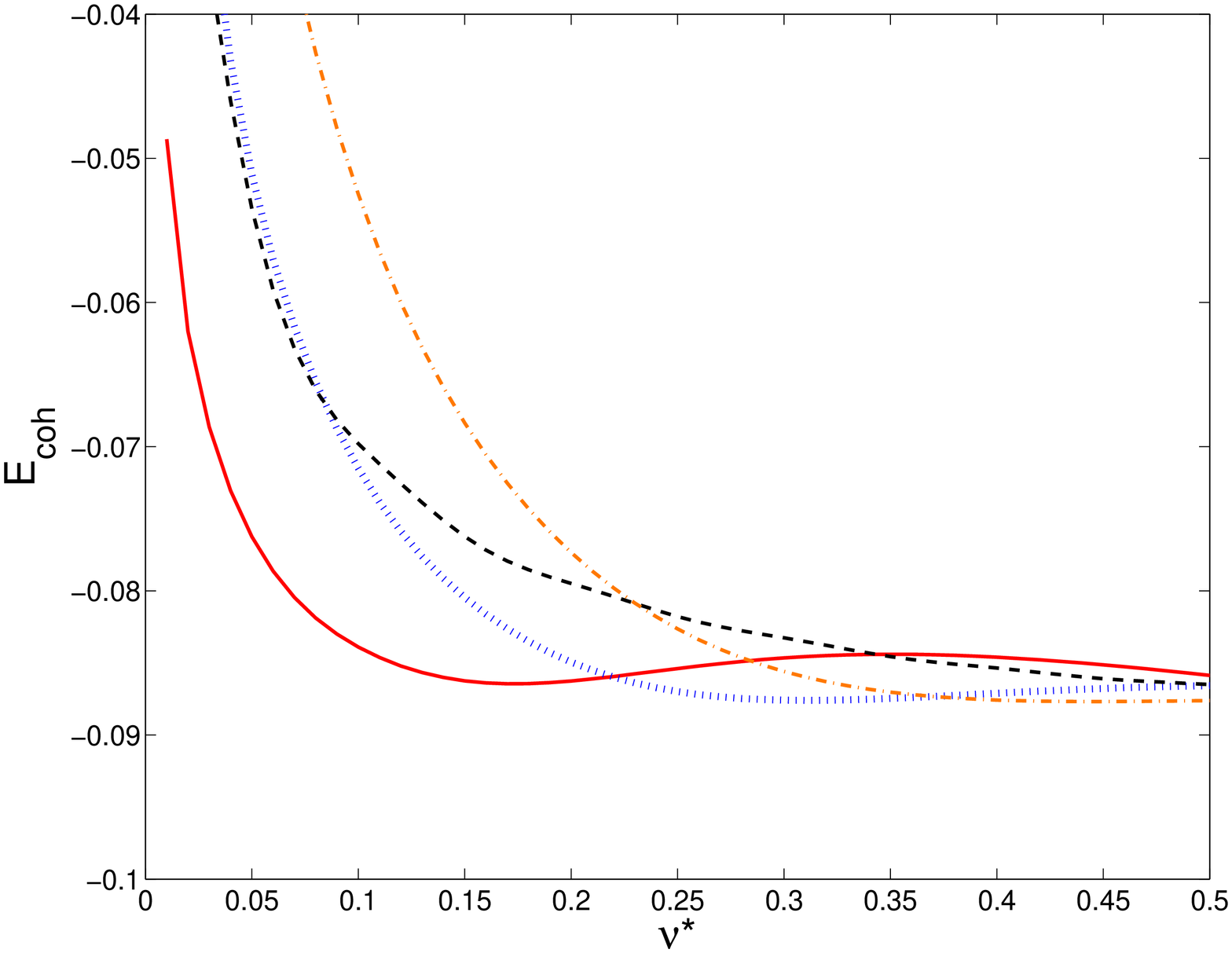}
\caption{(Color online) Cohesive energies of the Wigner and bubble crystals
in the $n=2$ LL in units of $e^2/\epsilon\ell$, with the interaction potential
$V_{HF}$ resulting from the effective Coulomb potential in Eq.~(\ref{newV}). The solid line is for the
Wigner crystal, the dotted line is for the two-electron bubble phase, the dash-dotted line is for
the $M=3$ BC, and the dashed line is for the stripe phase. The upper panel is for $\lambda=0$,
and the lower panel is for $\lambda=1$.}
\label{n2screened}
\end{figure}

\begin{figure}[t]
\centering
\includegraphics[angle=-90,totalheight=4.7cm,width=8.09cm,viewport=5 5 555 725,clip]{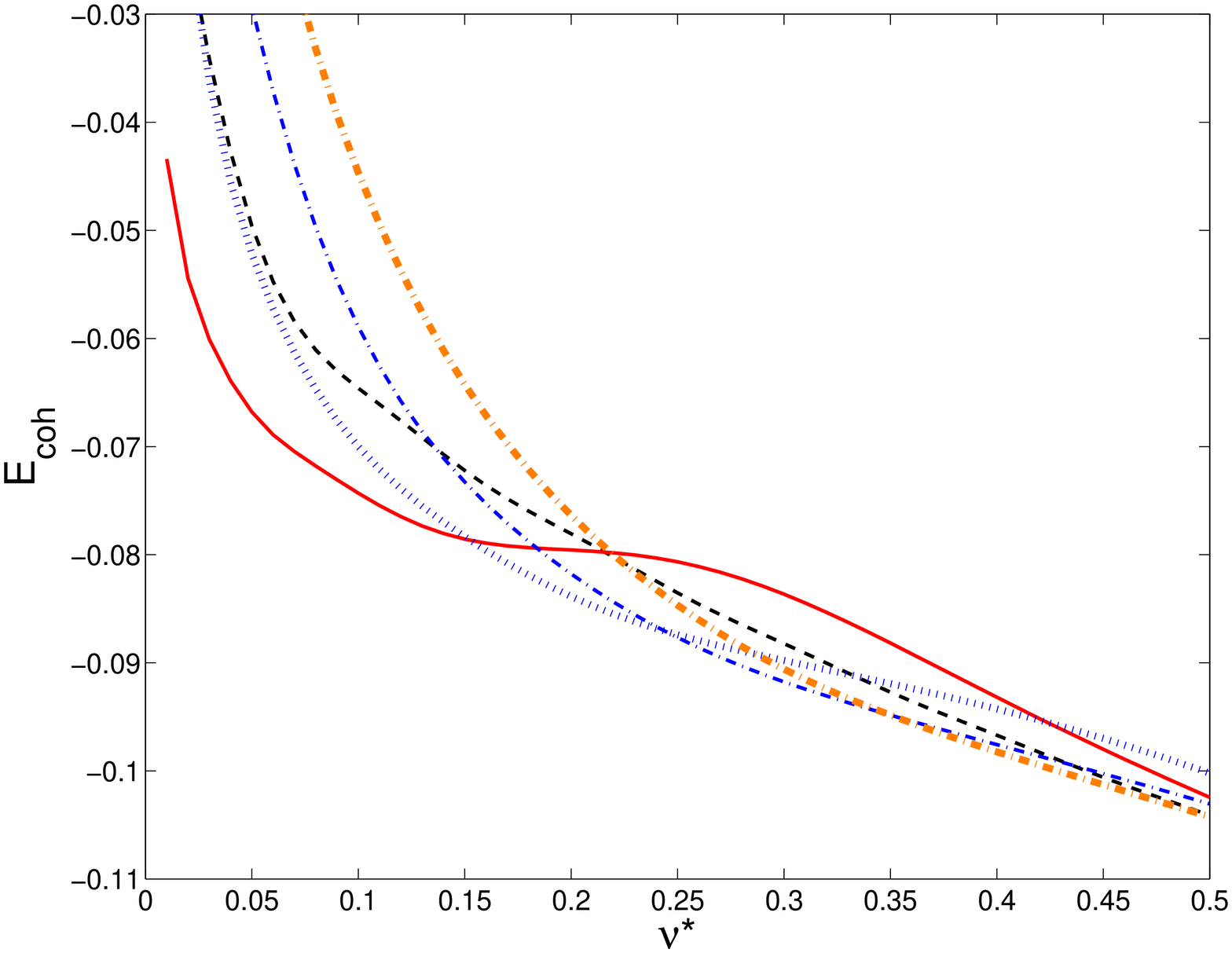}
\includegraphics[angle=-90,totalheight=4.7cm,width=8.09cm,viewport=5 5 555 725,clip]{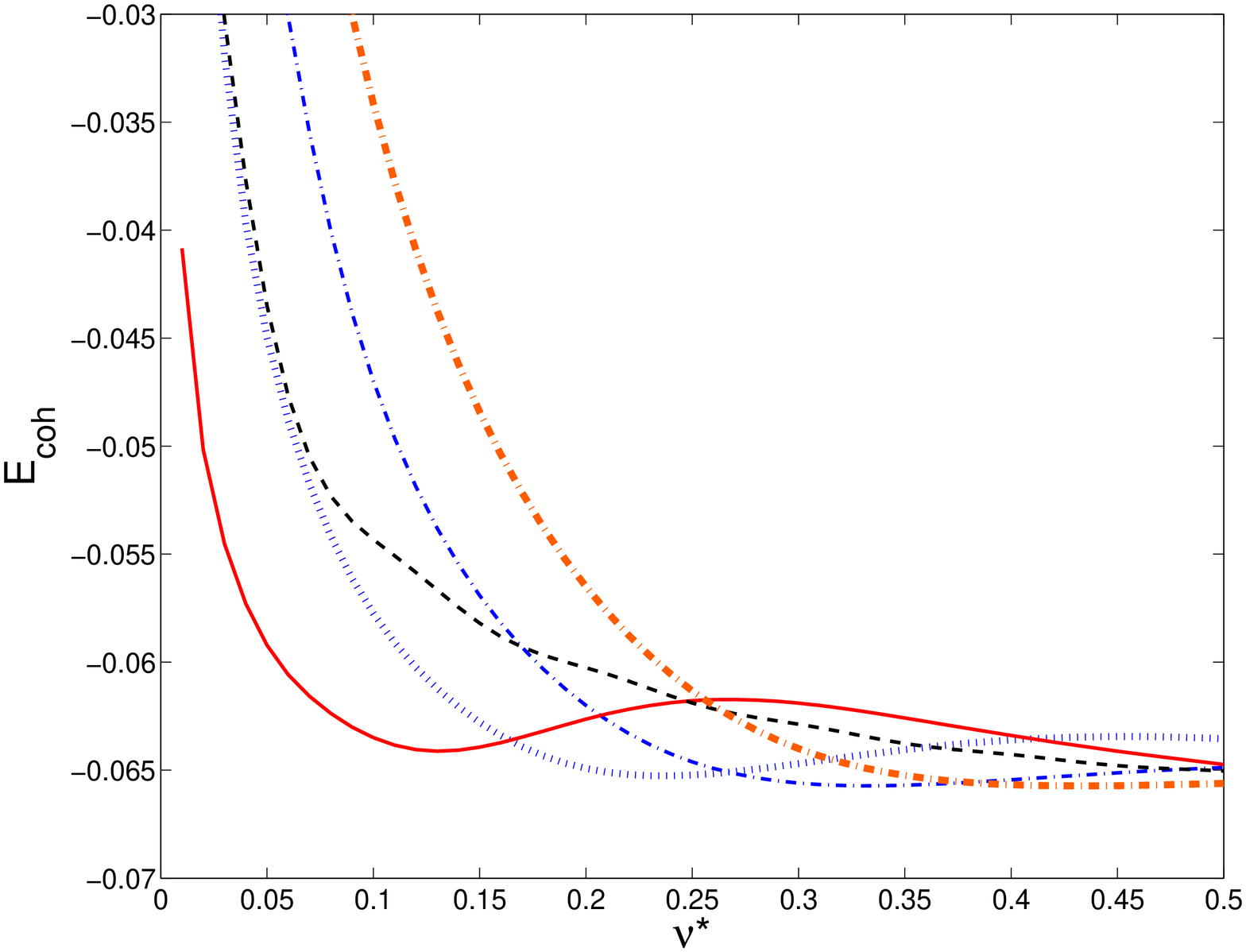}
\caption{(Color online) Same as in Fig.~\ref{n2screened}, but for Landau level $n=3$. The thick dash-dotted line is
for the bubble state with $M=4$.}
\label{n3screened}
\end{figure}

The form of the effective Coulomb interaction of Eq. (\ref{newV})
makes it difficult to find an analytic expression for the Fock part
of the interaction potential, Eq. (\ref{VF}). To evaluate the
cohesive energy, the ${\bf q}$ integration in this last equation is
performed numerically. In Figs.~(\ref{n2screened}) and
(\ref{n3screened}), we plot the cohesive energies of the Wigner
crystal, bubble phases and stripe states in the second ($n=2$) and
third ($n=3$) Landau levels, respectively. In each of these two
figures, the upper panel corresponds to $\lambda=0$, while the lower
panel corresponds to $\lambda=1$. The screening by lower LLs has the
most drastic effect on the cohesive energy, shifting it upwards by
about $50$ percent of its bare value, and finite sample thickness
also tends to shift the cohesive energies up, although in a less
pronounced fashion. It is noteworthy that these shifts in the
cohesive energy curves do not alter the overall phase diagram. 
As can be seen from Fig. \ref{n2screened}, the transitional values
$\nu^*_{1,2}$ from WC to the 2e bubble phase, and $\nu^*_{2,3}$ from
$M=2$ to $M=3$, are slightly shifted downwards by screening alone
(we find $\nu^*_{1,2}\simeq 0.2$ and  $\nu^*_{1,2}\simeq 0.35$ with screening, 
while $\nu^*_{1,2}\simeq 0.22$ and $\nu^*_{2,3}\simeq 0.37$ without screening).
Finite thickness effects on the other hand shift transitional filling factors 
upwards, restoring them to values that are very close to their bare values.
In all cases, these shifts do not affect the overall phase diagram, which
remains the same as in the ideal case of zero sample thickness
and no screening by lower Landau levels.

\noindent

We have also calculated the shear moduli of the Wigner and bubble
crystals in presence of screening and taking finite thickness
effects into account. As we mentioned above, in the presence of
screening, it becomes difficult to obtain an analytic expression for
the Fock part of the HF interaction potential, and the method of
Sec. \ref{ElasticModuli} becomes impractical. A more efficient way
to extract the elastic coefficients of that case consists in finding
the cohesive energy of a distorted crystal for a given uniform
deformation. The elastic energy will then be given by the excess
energy of the deformed crystal with respect to the reference
undistorted state. For two-dimensional uniform deformations, such
that the displacement vector ${\bf u}({\bf r})$ is given by
\begin{equation}
u_\alpha = u_{\alpha,\beta} x_\beta \,,
\end{equation}
with constant coefficients $u_{\alpha,\beta}$, it can be
shown\cite{Miranovic2001} that the reciprocal lattice also
experiences a homogeneous deformation, with the deformed reciprocal
lattice vectors $\{{\tilde{\bf Q}}\}$ given in terms of the original
ones $\{{\bf Q}\}$, to first order in the small displacements
$\{u\}$, by:
\begin{equation}
\tilde{Q}_\alpha = Q_\alpha - u_{\beta,\alpha} Q_\beta.
\label{tildeQ}
\end{equation}
To find a given elastic modulus, we calculate the HF energy of the
corresponding distorted state (with distortion amplitude $u_0$), and
extract the elastic constant from the excess energy
$E_{HF}(u_0)-E_{HF}(0)$. For example, to extract the shear modulus
$c_{66}$ we calculate the HF energy $E_{HF}(u_0)$ of the distorted
crystal using the following reciprocal lattice vectors:
\begin{subequations}
\begin{eqnarray}
\tilde{Q}_x & = & Q_x - u_0 Q_y\,,
\\
\tilde{Q}_y & = & Q_y\,,
\label{tildeQC66x}
\end{eqnarray}
\end{subequations}
corresponding to the shear deformation polarized along $y$ such that
${\bf u}({\bf r}) = u_0 \, x \,\hat{\bf y}$.
The shear modulus (in units of {\em energy/particle}) is then given by:
\begin{equation}
c_{66} = \frac{2}{u_0^2}\,\big[E_{HF}(u_0)-E_{HF}(0)\big].
\end{equation}
Note that this procedure allows us to extract the elastic moduli
{\em without} Taylor expanding in the small displacement vectors
$\{{\bf u}\}$, which allows us to better handle anharmonic effects
which might be important in case quantum fluctuations happen to be
large\cite{Fisher1982,Cote1990} (which might be the case with the softer lattices
in presence of screening by lower LLs). Note also that, since cohesive
energy calculations involve reciprocal lattice sums that avoid
summing over the (diverging) contribution of the ${\bf Q}=0$ Hartree
term, one should add to the results for the compression modulus
$c_{11}$ obtained by the above method a term $\frac{1}{A_c}U_H(q)$
to obtain the correct compression moduli of the electronic solid.

\begin{figure}[t]
\centering
\includegraphics[angle=-90,totalheight=4.7cm,width=8.09cm,viewport=5 0 560 720,clip]{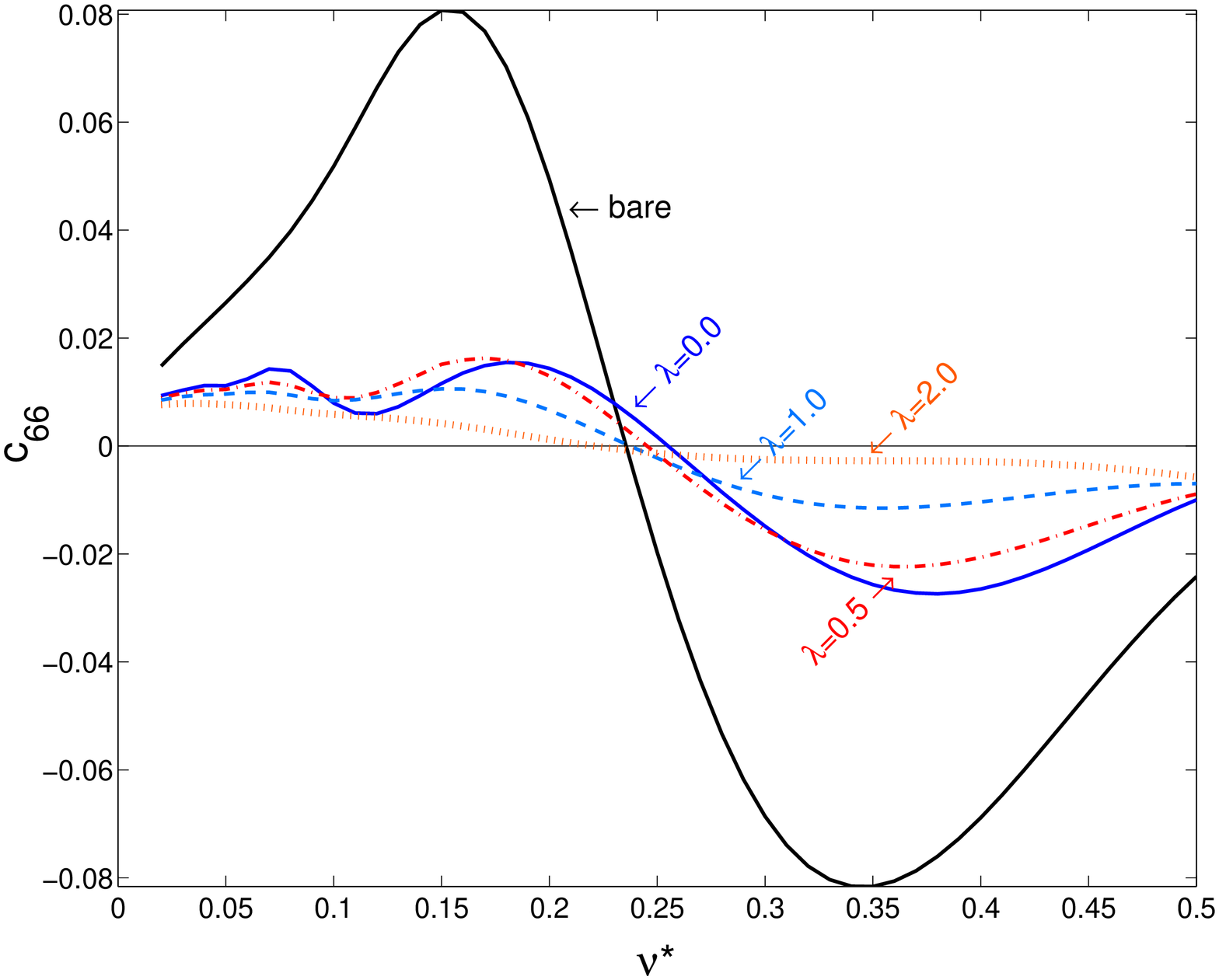}
\includegraphics[angle=-90,totalheight=4.7cm,width=8.09cm,viewport=5 5 560 730,clip]{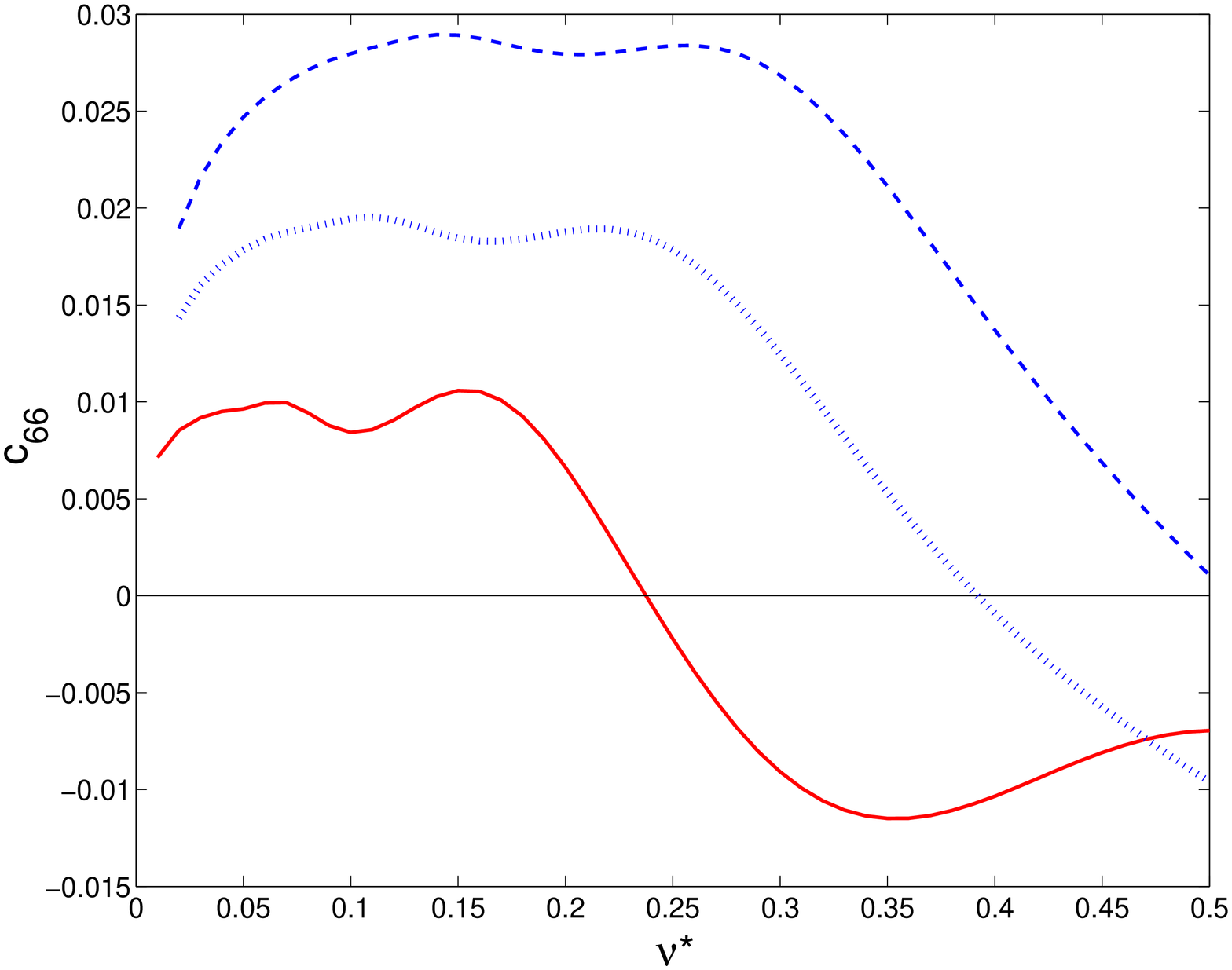}
\caption{(Color online) Upper panel: Shear moduli (in units of $e^2/\epsilon\ell$)
for the WC (solid lines) in the $n=2$ LL with the interaction
potential $V_{HF}$ resulting from the effective Coulomb potential in
Eq.~(\ref{newV}). The succession of curves is for $\lambda =
0,\,0.5,\,1$ and $2$. Lower panel: shear moduli of the WC and for
the 2 and 3 electron bubble solids (from bottom to top) in presence
of screening by lower LLs and with $\lambda=1$.} \label{c66screened}
\end{figure}

\begin{figure}[t]
\centering
\includegraphics[angle=-90,totalheight=4.7cm,width=8.09cm,viewport=5 15 560 795,clip]{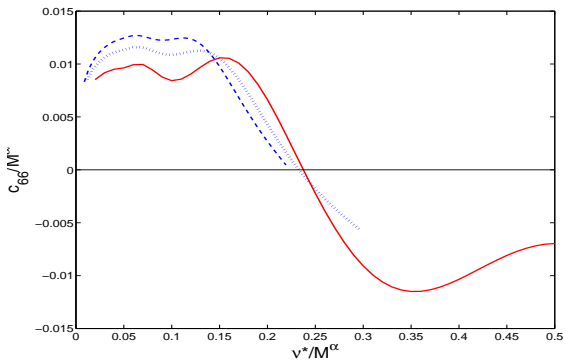}
\caption[]{
Universal scaling of the shear modulus for the Wigner and bubble crystals in the $n=2$ Landau level in presence
of screening and finite thickness effects. Here we used the value $\lambda=1$ for the screening strength, and
the value $\alpha\simeq 0.75$ for the scaling exponent.
}
\label{Fig:scaling2}
\end{figure}

In order to test the validity of the above procedure, we have
calculated the shear moduli of the Wigner and $M=2$ bubble crystals
in the $n=2$ Landau level. As can be seen in Fig. \ref{c66screened}
(upper panel), we obtain the same shear modulus for the WC as in
Sec. \ref{ElasticModuli} where we used a Taylor expansion of the
interaction energy in the small displacements $\{{\bf u}({\bf
r})\}$, which is a consistency check on our calculations. In
Fig.~\ref{c66screened}, we also show the results we obtain for the
shear modulus of the WC in presence of screening and for various
values of the parameter $\lambda$. It is seen that screening by
lower LLs alone ({\em i.e.}, with $\lambda=0$) considerably softens
the shear modulus with respect to the unscreened case, and higher
values of $\lambda$ lead to even smaller values of the shear modulus
of the WC. In the lower panel of Fig.~\ref{c66screened}, we plot the
shear moduli of the WC and of the 2 and 3 electron bubble states in
the $n=2$ LL in presence of screening and using $\lambda=1$. As it
can be seen, the shear moduli of these various types of crystals
are affected by the screening of the Coulomb interaction between
electrons in similar but nonequivalent ways, with the strongest effect impacting the
WC. This is quite understandable, given that a shear deformation of
a BC does {\em not} affect the part of the cohesive energy coming
from electrons inside the bubbles. It is to be noted also that
the universal scaling of Eq. (\ref{Eq:scaling}) holds only approximatively
in the screened case, as shown in Fig. \ref{Fig:scaling2}, where we used
$\alpha = \beta \simeq 0.75$. We shall discuss some of the
implications of our results in Sec. \ref{Consequence}.

\section{Link with experiment: Consequence for the microwave conductivity of two dimensional bubble crystals}
\label{Consequence}

We now would like to discuss the experimental implications of our
results. Our investigation of the dynamics of bubble solids stemmed
primarily from a desire to understand recent microwave conductivity
experiments by Lewis {\em et al.}\cite{Lewis2004} in the second
($n=2$) Landau level, where the appearance of a second resonance
peak around $\nu^*\sim 0.16$ was attributed to the formation of a
bubble phase, which coexists with a Wigner crystal over a range of
values of $\nu^*$. In what follows, we shall attempt a qualitative
analysis of the experimental results of Lewis {\em et al.}\cite{Lewis2004}
in light of the results of the elasticity theory presented in this paper combined
with the theoretical predictions of Refs. \onlinecite{Fogler2000,Chitra2001}.

According to Chitra et {\em al.},\cite{Chitra2001} the resonance
frequency due to pinning of a two dimensional Wigner crystal in a
strong magnetic field is given by
\begin{equation}
\omega_p \simeq \frac{\Sigma}{\rho_m \omega_c},
\label{omegap}
\end{equation}
where $\rho_m=m/({\pi}a^2)$ is the mass density, $\omega_c=eB/m$ is the cyclotron frequency, and where
$\Sigma$ is given by (we neglect an overall numerical constant of order unity):
\begin{equation}
\Sigma \simeq \frac{c_{66}}{R_a^2}\big(\frac{a}{\xi_0}\big)^6.
\label{Sigma}
\end{equation}
In the above expression, $\xi_0$ is the greatest of $\ell$ and the correlation length $\xi_d$ of the
disorder potential.
On the other hand, $R_a$ is the length scale at which electron displacements 
become of order $a$, and is given by (here $\Delta$ is the variance of the random pinning potential, 
whose distribution was assumed by Chitra {\em et al.} to be Gaussian and of short range):
\begin{equation}
R_a = \frac{\pi c_{66}a^4}{\sqrt{\Delta}}. 
\label{Ra}
\end{equation}
Using Eq.~(\ref{Ra}) into Eq.~(\ref{Sigma}), we obtain:
\begin{equation}
\Sigma \sim \frac{\Delta}{c_{66}\xi_0^6a^2}, 
\label{Sigma2}
\end{equation}
and hence
\begin{equation}
\omega_p \simeq \frac{\Delta}{\rho_m\omega_c c_{66}\xi_0^6a^2}. 
\label{Sigma3}
\end{equation}
Using the fact that $\rho_m=m/\pi a^2$, we see that $\rho_m a^2 = m$, and hence:
\begin{equation}
\omega_p(\nu^*) \sim \frac{\Delta}{m\omega_c \xi_0^6 c_{66}}.
\label{omegap2prime}
\end{equation}
In the above equation, $m$ is the mass of electrons within a unit lattice cell, and hence
$m=M m^*$, where $m^*$ is the effective electron mass in the host semiconductor.
We thus obtain:
\begin{equation}
\omega_p(\nu^*) \sim \frac{\Delta}{Mm^*\omega_c \xi_0^6 c_{66}}.
\label{omegap2}
\end{equation}
The above result implies that the $\nu^*$ dependence of the resonance frequency $\omega_p$ arises
mainly from the dependence of the effective shear modulus $c_{66}$ on the
partial filling factor $\nu^*$ (the dependence of $\xi_0$ and $\omega_c$ on $\nu^*$ being
rather weak in Landau levels of index $n\ge 2$). We thus see that knowledge of the variation of 
$c_{66}$ vs. $\nu^*$ in a given Landau level and for a given bubble crystal should allow 
us to easily infer the variation of $\omega_p$ vs. $\nu^*$. Conversely, analysis 
of experimental data of $\omega_p$ vs. $\nu^*$ using the above expression
(and our Hartree-Fock results for $c_{66}(\nu^*)$) should allow us to identify 
Wigner crystal and bubble phases, and transitions between these phases.

We now turn our attention to the experimental results of Ref. \onlinecite{Lewis2004}, 
and more specifically to Fig.~3 in this last reference, where the 
authors plot experimental resonance frequencies of the two-dimensional 
electron system in the $n=2$ Landau level as a function of the filling 
factor $\nu$. At small $\nu^*$, there is only one resonance peak, 
whose frequency decreases with increasing $\nu^*$, with a minimum 
around $\nu^*\simeq 0.19$. The frequency of the above peak then slightly 
increases, until the peak disappears around $\nu^*\simeq 0.27$. 
This behavior is in qualitative agreement with the behavior of a 
Wigner crystal if we use our shear modulus of Fig.~\ref{figc66} 
(Wigner crystal in the $n=2$ LL) in conjunction with Eq.~(\ref{omegap2}).

Figure 3 of Lewis {\em et al.} also shows the peak frequency of a
second resonance peak which appears at $\nu^*\ge 0.16$. The $\nu^*$
dependence of the frequency of this second resonance peak, however, does not seem 
to be fully consistent with Eq.~(\ref{omegap2}) above and with the shear 
moduli of the $M=2$ bubble phase plotted in Fig.~\ref{figc66}. 
Indeed, in the range $0.23\leq \nu^*\leq 0.35$, from the decreasing
behavior of the shear modulus of the bubble phase in Fig. \ref{figc66}, we
expect an increasing behavior of the peak frequency $\omega_p$ vs. $\nu^*$.
The experimental result of Ref. \onlinecite{Lewis2004} show a decreasing
peak frequency in this range of filling factors, in disagreement with
the theoretical considerations above. Equally important is the fact that 
Eq. (\ref{omegap2}) predicts a factor of $1/M$ difference between the frequencies 
of different bubble states, while experimentally it seems that a factor 
$1/\sqrt{M}$ is observed instead.

It is to be noted that, in their recent replica study of bubble phases pinned by
random disorder, the authors of Ref. \onlinecite{Cote2005} claim to have been able
to successfully describe the $\nu^*$ dependence of the resonance peak of both
Wigner and bubble crystals, except on the region between $\nu^*=0.16$
and $\nu^*=0.28$ where the two phases are assumed to coexist. We therefore
may speculate that the disagreement we find in the present study
may be due to a possible breakdown of Eq. (\ref{omegap}), which in fact
is derived by Chitra {\em et al.}\cite{Chitra2001} by linearizing nonlinear
self-consistent replica calculation, and that a careful and exhaustive
treatment such as the one in Ref. \onlinecite{Cote2005}, where the full nonlinear
replica equations are numerically solved in a self-consistent way,
can lead to better agreement between theoretical predictions and experimental 
observations.

We now briefly comment on the effect of screening by lower LLs and of finite
sample thickness (as modeled by the parameter $\lambda$ of the Zhang-Das Sarma potential
of the previous Section) on the microwave response. Since we have shown that both effects tend
to reduce the shear modulus with respect to the bare case, we conclude that the resonance peak
will tend to shift toward higher frequencies as the sample thickness or Landau level index is increased,
an effect that may be testable experimentally.
More importantly, in presence of screening and for finite thickness samples,
we find that there is a finite gap between the shear moduli of the Wigner and 
bubble crystals, and that gap might accentuate the finite gap observed in the
coexistence region between the pinning frequencies of the Wigner and the 2e bubble crystals,
making it more pronounced. \cite{Lewis2004}
In fact, it would be interesting to do a full self-consistent calculation, like the one done
in ref. \onlinecite{Cote2005}, in presence of screening and finite thickness effects,
and see whether the experimentally observed gap between the pinning frequencies in the coexistence region
can be reproduced theoretically.

\section{Conclusions}
\label{Discussion}

In conclusion, in this paper we have calculated the cohesive energies,
shear moduli, and normal (magnetophonon and magnetoplasmon) modes of the Wigner
and bubble crystals in higher LLs. Going beyond previous treatments,
we have studied the effects of screening by lower Landau levels as
well as the case of finite sample thickness. We found that both effects
reduce the cohesive energies and elastic moduli, with the screening
by lower Landau levels having the most pronounced effect. The transitional
values of the filling factor as well as the overall nature of the phase
diagram remain, however, unchanged by both effects. We have also
examined the electromagnetic response of the Wigner and bubble crystals,
and have noticed that, while the $\nu^*$ dependence of the shear modulus 
of the WC, when used in standard theories of the electromagnetic response 
of electronic crystals, is in good qualitative agreement with experimental data, 
the shear modulus of the 2e BC has a $\nu^*$ dependence which, in contrast, 
is in disagreement with the experimentally observed behavior. 
This led us to suggest that a full numerical solution of the non-linear self-consistent
replica equations\cite{Cote2005} may be needed in order to understand the electromagnetic
response of bubble crystals, by contrast to Wigner crystals for which
a linearized estimate\cite{Chitra2001} of the resonance frequency may be adequate.

Before closing, we briefly comment on an issue that has not been
previously addressed in the literature, and that is the issue of
gauge invariance. In principle, all physical observables are
gauge-independent. However, HF being an approximate ({\em i.e.} not
an exact) theory, different HF derivations done in different gauges
will not necessarily give the same results. As a check to our
calculations of the cohesive energies of the Wigner and bubble
crystals, and to make sure that the same qualitative phase diagram
that we obtained above in the Landau gauge is not affected by a
different choice of gauge, we have rederived the HF interaction
potentials and the phase diagram of the 2D electron system in the
symmetric gauge, using a HF approach similar to the one
used a long time ago by Maki and Zotos\cite{Maki1983} to study the
properties of the two-dimensional Wigner crystal in the lowest
Landau level. In Appendix \ref{appWvfct}, we derive expressions for
the interaction potential $U_{mm'}(r)$ between electrons of angular
momenta $m$ and $m'$, with $m,m'=0,1,2$ by finding the average Coulomb
energy between symmetric-gauge electronic wavefunctions in the
$n=2$ Landau level. These interaction potentials have the same
qualitative behavior as the potentials derived within the Landau gauge HF
approach of the text, and lead to very similar cohesive energies as
the ones predicted above, although with slightly different numerics.

\acknowledgments

The authors acknowledge discussions with R.M. Lewis, L.W. Engel, K. Yang,
M.M. Fogler, C. Doiron, R. C\^ot\'e and H.A. Fertig.
This work has been supported by the National High Magnetic Field Laboratory In
House Research Program.

\appendix

\section{Projected densities and HF potentials}
\label{appRhos}

In order for our paper to be self-contained, in this Appendix we give the explicit expressions of the
projected densities and Hartree-Fock potentials that we used in deriving our cohesive energies. From the
usual expressions of the noninteracting wavefunctions $\varphi_{nm}({\bf r})$ (see Eq.~(\ref{wvfcts}),
and the definitions (\ref{rho(q)})-(\ref{tilderho}), we obtain that the projected densities
$\tilde{\rho}_0({\bf q})$ and $\tilde{\rho}_1({\bf q})$ are given, both in $n=2$ and $n=3$, by
(throughout this Appendix, $q$ stands for the dimensionless quantity $q\ell$)
\begin{subequations}
\begin{gather}
\tilde{\rho}_0({\bf q}) = e^{-q^2/4},\\
\tilde{\rho}_1({\bf q})=(1-\frac{1}{2}q^2)e^{-q^2/4}.
\end{gather}
\end{subequations}
On the other hand, performing the integrals in Eqs.~(\ref{VH})-(\ref{VF}), we obtain the following Hartree
and Fock potentials in LL $n=2$:
\begin{widetext}
\begin{subequations}
\begin{align}
V_H(q)&=\Big(\frac{2\pi e^2\ell}{\epsilon}\Big)\frac{e^{-q^2/2}}{64q}\big(8 - 8q^2 + q^4\big)^2, 
\\
V_F(q)&=-\Big(\frac{2\pi e^2\ell}{\epsilon}\Big)
\frac{e^{-q^2/4}}{128}\sqrt{\frac{\pi}{2}}\Big[(82-52q^2+44q^4-10q^6+q^8\big)
\mbox{I}_0(\frac{q^2}{4})-q^4\big(30-8q^2+q^4\big)\mbox{I}_1(\frac{q^2}{4})\Big],
\end{align}
\end{subequations}
while for LL $n=3$ the Hartree and Fock potentials are given by:
\begin{subequations}
\begin{align}
V_H({\bf q})&=\Big(\frac{2\pi e^2\ell}{\epsilon}\Big)\,\frac{e^{-q^2/2}}{2304q}\big(48-72q^2+18q^4-q^6\big)^2,
\\
V_F({\bf q})&=-\Big(\frac{2\pi e^2\ell}{\epsilon}\Big)\frac{e^{-q^2/4}}{4608}\sqrt{\frac{\pi}{2}}\Big[\big(2646-2430q^2+2889q^4-1236q^6+270q^8
-26q^{10}+q^{12}\big)\mbox{I}_0(\frac{q^2}{4})\notag\\
&-q^4\big(1539-828q^2+224q^4-24q^6+q^8\big)\mbox{I}_1(\frac{q^2}{4})\Big].
\end{align}
\end{subequations}
\end{widetext}

\section{Cohesive energy of the stripe phase}
\label{appStripe}

For completeness, in this Appendix, we briefly review the cohesive
energy of the stripe phase of the two-dimensional electron system.
Following Fogler {\em et al.}\cite{Fogler1996} and Goerbig {\em et
al}.\cite{Goerbig2004}, we write the cohesive energy per particle
$E_{coh}$ in the form
\begin{equation}
E_{coh}= \frac{n_B}{2\nu^*}\sum_{\bf q}V_{HF}(q) |\Delta(q)|^2
\label{cohstr}
\end{equation}
where $n_B=1/(2\pi\ell^2)$ is the flux density, and where
$\Delta({\bf q})=\rho({\bf q})/(n_B A)$ is the Fourier
transform of the local guiding center filling factor $\nu^*({\bf
r})$ ($A$ is the area of the sample). For a stripe state,
$\nu^*({\bf r})=\sum_i\Theta(|x-x_i|-a/2)$, with $a$ the width of a
given stripe, and $x_i$ the location of the center of the $i$-th
stripe. If we denote by $L_s$ the stripe periodicity (such that
$x_i=iL_s$), and note that\cite{Fogler1996,Goerbig2004}
$\nu^*=a/L_s$, then we obtain after a few manipulations that the
cohesive energy per particle of Eq.~(\ref{cohstr}) is given by
\begin{equation}
E_{coh}= \frac{n_B}{2\pi^2\nu^*}{\sum_{n}}'V_{HF}\Bigg(\frac{2\pi
n}{L_s}\Bigg) \frac{\sin^2(\pi\nu^* n)}{n^2}, \label{cohstr2}
\end{equation}
where the prime on the sum sign indicates that the (diverging) Hartree contribution for $n=0$ is excluded
from the summation (the $n=0$ Fock term being included).
The above expression is then minimized with respect to $L_s$, 
with the optimal stripe periodicity found\cite{Goerbig2004} to be approximately
given by $L_s^*\simeq 2.76\ell\sqrt{2n+1}$ in $n=2$ and $L_s^*\simeq 2.74\ell\sqrt{2n+1}$ in $n=3$. 
The cohesive energy in the stripe phase is then given by the optimal
value $E_{coh}(L_s^*)$.

\section{Gauge dependence of the cohesive energies: Hartree-Fock approach in the symmetric gauge}
\label{appWvfct}

\begin{figure}[t]
\centering
\includegraphics[width=8.09cm, height=5cm]{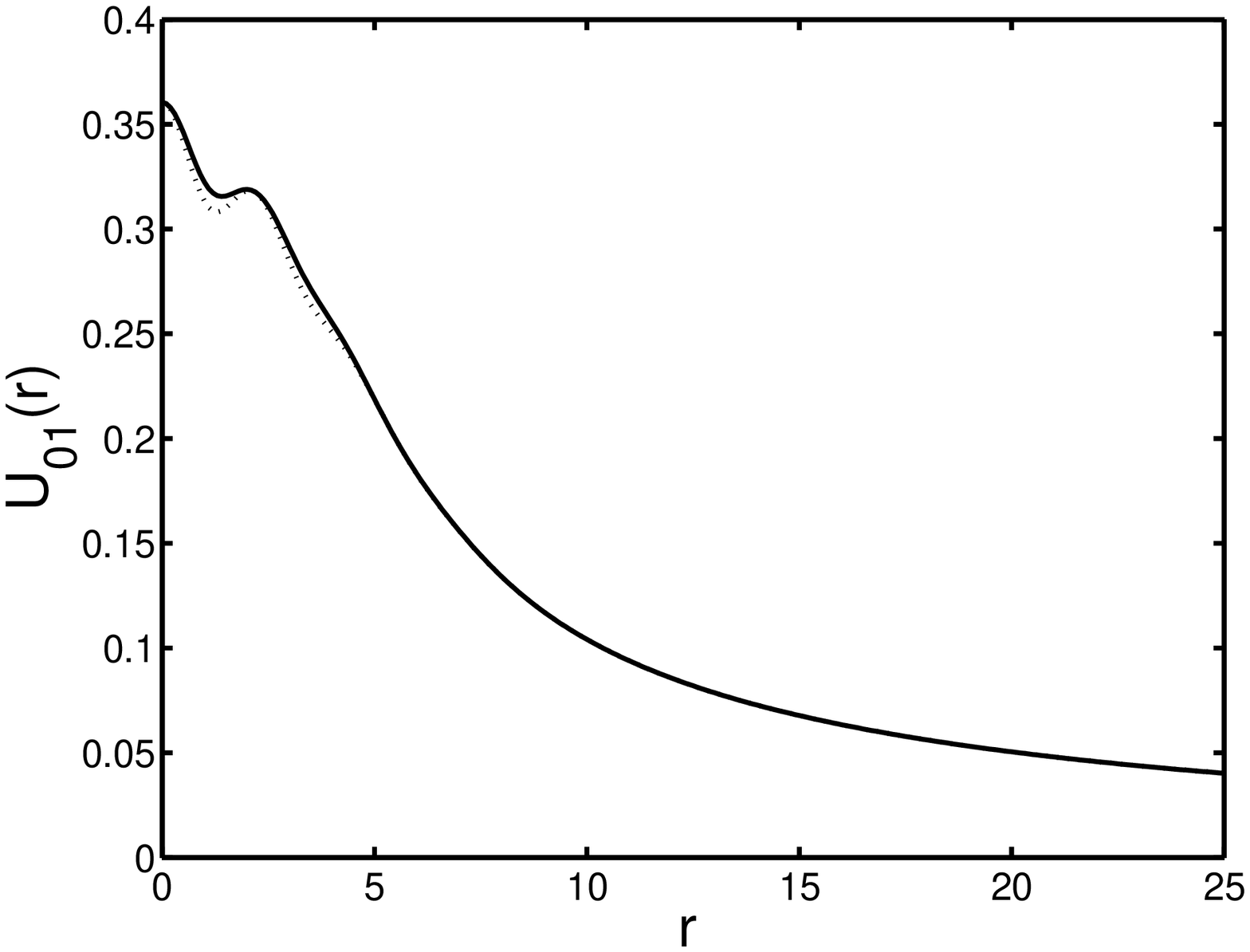}
\includegraphics[width=8.09cm, height=5cm]{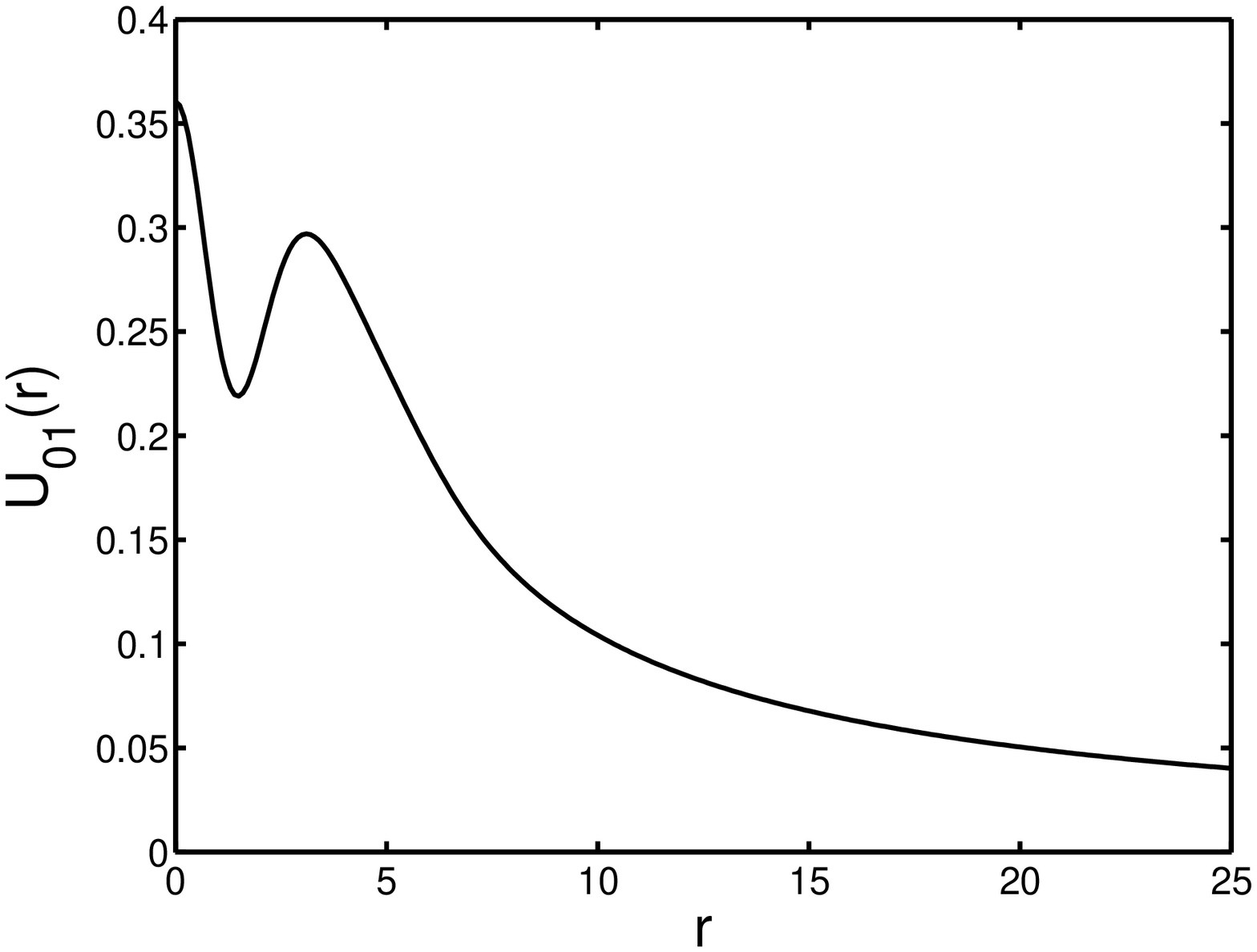}
\caption{Upper panel: Interaction potential $U_{01}(r)$ (in units of $e^2/\ell$) vs. $r$ (in units of $\ell$) as obtained
from the HF approach in the symmetric gauge, Eqs. (\ref{N01})-(\ref{D01}), in the $n=2$ Landau level.
The dotted line corresponds to the numerator $N_{01}(r)$ alone.
Lower panel: Effective interaction potential $U_{01}(r)$ in the Landau gauge.}
\label{figU01wvfct}
\end{figure}

\begin{figure}[t]
\centering
\includegraphics[width=8.09cm, height=5cm]{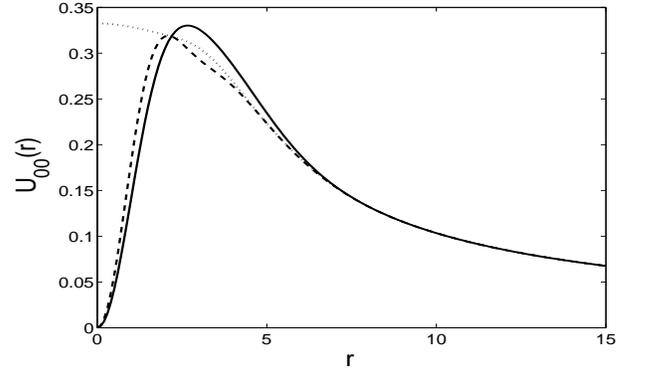}
\caption{Interaction potential $U_{00}$ (in units of $e^2/\ell$) vs.
$r$ (in units of $\ell$) in the $n=2$ Landau level. The solid line
is the Landau gauge result, while the dashed line is the result for
$N_{00}(r)$ in the symmetric
gauge, Eq.~(\ref{N00}). The dotted line is the whole function
$U_{00}(r)=N_{00}(r)/D_{00}(r)$. (Note that this last function has
no physical meaning at small separations, {\em i.e.} for $r\leq
\ell$, due to the Pauli exclusion principle for Fermions.)}
\label{figU00wvfct}
\end{figure}

In this Appendix, we want to confirm the qualitative features of the
phase diagram of electrons in higher LLs derived in the text using a
different gauge for the applied magnetic field, namely the symmetric gauge
${\bf A}=\frac{B}{2}(-y\hat{\bf x}+x\hat{\bf y})$. To this end, we
shall use a HF approach which is similar in spirit to the
approach of Maki and Zotos\cite{Maki1983} to derive the effective
interaction potentials $U_{mm'}(r)$ in the second ($n=2$) Landau
level. Let us consider two electrons with angular momenta $m_1$ and
$m_2$ and with guiding centers located at lattice sites ${\bf R}_1$
and ${\bf R}_2$, respectively. We shall use for the wave function of
the two-bubble system the following antisymmetric combination
\begin{eqnarray}
\Psi({\bf r}_1,{\bf r}_2) &=& \frac{1}{\sqrt{2}}\big(\varphi_{nm_1{\bf R}_1}({\bf r}_1)
\varphi_{nm_2{\bf R}_2}({\bf r}_2)
\notag\\
&-&\varphi_{nm_1{\bf R}_1}({\bf r}_2)\varphi_{nm_2{\bf R}_2}({\bf r}_1)\big),
\end{eqnarray}
where $\varphi_{nm{\bf R}}({\bf r})$ is the symmetric gauge wavefunction with angular momentum $m$ which is located around
the lattice site ${\bf R}_k$, $\varphi_{nm\bf R}({\bf r})=\varphi_{nm}({\bf r}-{\bf R})$, with
\begin{eqnarray}
\varphi_{nm}({\bf r}) &=& C_{nm}\big(\frac{r}{\ell}\big)^{|n-m|}e^{i(n-m)\theta}e^{-r^2/4\ell^2}
\notag\\
&\times& L^{|n-m|}_{(n+m-|n-m|)/2}\big(\frac{r^2}{2\ell^2}\big),
\label{wvfcts}
\end{eqnarray}
where we denote by $\theta$ the polar angle of vector ${\bf r}$, {\em i.e.}
${\bf r}=(r\cos\theta,r\sin\theta)$, and
where the normalization constant $C_{nm}$ is given by:
\begin{gather}
C_{nm} = \frac{1}{\ell}\begin{cases}
\sqrt{\frac{2^{n - m}n!}{2\pi m!}} & \mbox{if $m\ge n$,}\\
{ } & { }\\
\sqrt{\frac{2^{m - n}m!}{2\pi n!}} & \mbox{if $n\ge m$.}
\end{cases}
\end{gather}
The average potential energy for this state is given by (we here for simplicity set $\epsilon=1$):
\begin{equation}
U_{m_1m_2}({\bf R}_1,{\bf R}_2) = \frac{\langle \Psi|\frac{e^2}{|{\bf r}_1-{\bf r}_2|}|\Psi \rangle}
{\langle \Psi | \Psi \rangle}.
\label{poten}
\end{equation}
Note that the above expression is valid for arbitrary ${\bf R}_1$ and ${\bf R}_2$ only if $m_1\neq m_2$.
In the special case $m_1=m_2$, because of the Pauli exclusion principle for Fermions
the above expression for $U_{m_1m_2}$ only makes sense
at separations $|{\bf R}_1-{\bf R}_2|$ larger than the typical extension of the wavefunctions
$\varphi_{nm}({\bf r})$, which is of order $\ell$.

Now, the numerator of the above expression is given by 
(to simplify the notation, in the rest of this Appendix we shall drop
the LL index $n$ from the $\varphi_{nm}$ wavefunctions):
\begin{widetext}
\begin{gather}
N_{m_1,m_2}({\bf R}_1,{\bf R}_2)=e^2\int \frac{d{\bf r}_1d{\bf r}_2}{|{\bf r}_1-{\bf r}_2|}
\big[|\varphi_{m_1,{\bf R}_1}({\bf r}_1)|^2|\varphi_{m_2,{\bf R}_2}({\bf r}_2)|^2-\varphi_{m_1,{\bf R}_1}^*
({\bf r}_1)\varphi_{m_2,{\bf R}_2}^*({\bf r}_2)\varphi_{m_2,{\bf R}_2}({\bf r}_1)\varphi_{m_1,{\bf R}_1}
({\bf r}_2)\big].
\intertext{The denominator on the other hand is given by}
D_{m_1,m_2}({\bf R}_1,{\bf R}_2)=1-\int d{\bf r}_1 d{\bf r}_2\varphi_{m_1,{\bf R}_1}({\bf r}_1)
\varphi_{m_1,{\bf R}_1}^*({\bf r}_2)\varphi_{m_2,{\bf R}_2}({\bf r}_2)\varphi_{m_2,{\bf R}_2}^*({\bf r}_1)
=1-\big|\int d{\bf r}_1\varphi_{m_1,{\bf R}_i}({\bf r}_1)\varphi_{m_2,{\bf R}_j}^*({\bf r}_1)\big|^2.
\intertext{Now, using the fact that}
\varphi_{m\bf R}({\bf r})=\varphi_m({\bf r}-{\bf R})=\int_{\bf q}\varphi_{m}({\bf q})e^{-i{\bf q}
\cdot({\bf r}-{\bf R})},
\intertext{one can show, after a few manipulations, that the Hartree part of $N_{m_1m_2}$ is given by}
N_{m_1m_2}^H({\bf R}_i,{\bf R}_j)=\int_{\bf q}\frac{2\pi e^2}{q}\big(\int d{\bf r}_1|\varphi_{m_1}({\bf r}_1)|^2
e^{i{\bf q}\cdot{\bf r}_1}\big)
\big(\int d{\bf r}_2|\varphi_{m_2}({\bf r}_2)|^2 e^{-i{\bf q}\cdot{\bf r}_2}
\big)e^{i{\bf q}\cdot({\bf R}_i-{\bf R}_j)}. \label{NHartree}
\intertext{Similarly, the Fock part of $N_{m_1m_2}$ can be written in the form}
N_{m_1m_2}^F({\bf R}_i,{\bf R}_j)=-\int_{\bf q} \frac{2\pi e^2}{q}\big|\int d{\bf r}\varphi_{m_1}^*({\bf r})
\varphi_{m_2}({\bf r}+{\bf R}_i-{\bf R}_j)e^{i{\bf q}\cdot{\bf r}}\big|^2. \label{NFock}
\end{gather}
\end{widetext}
Finally, the denominator $D_{m_1m_2}$ can also be written in the form
\begin{equation}
D_{m_1m_2}({\bf R}_i,{\bf R}_j)=1-\big|\int d{\bf r}\varphi_{m_1}^*({\bf r})\varphi_{m_2}({\bf r}+{\bf R}_i
-{\bf R}_j)\big|^2.
\label{DFourier}
\end{equation}

\begin{figure}[t]
\centering
\includegraphics[width=8.09cm, height=5cm]{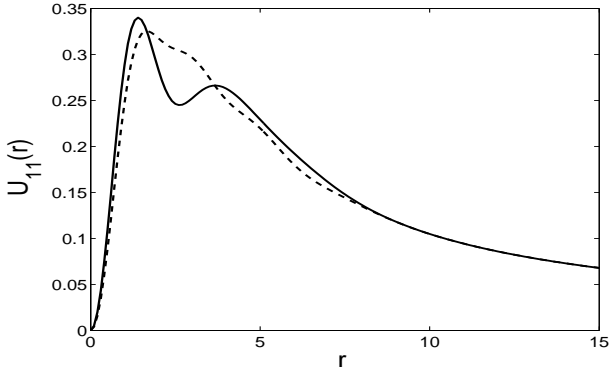}
\caption{Interaction potential $U_{11}$ (in units of $e^2/\ell$) vs. $r$ (in units of $\ell$) in the $n=2$
Landau level. The solid line is the Landau gauge result, while the dotted line is the 
symmetric gauge result for $N_{11}(r)$, Eq.~(\ref{N11}).}
\label{figU11wvfct}
\end{figure}

Eqs.~(\ref{NHartree})-(\ref{DFourier}) show that both Hartree and
Fock parts of $N_{m_1m_2}$ as well as the denominator $D_{m_1m_2}$ depend
only on the difference ${\bf R}_i-{\bf R}_j$ and are thus
translationally invariant as they should be. Using for the
$\varphi_{m,{\bf R}}$'s the wavefunctions for noninteracting
electrons in the $n$-th LL, Eq.~(\ref{wvfcts}), and performing the
integrations in Eqs.~(\ref{NHartree})-(\ref{DFourier}) (which is
most efficiently done in cartesian coordinates), we obtain that the
above potential energy, Eq.~(\ref{poten}), can be written in the
form:
\begin{equation}
U_{m_1m_2}({\bf R}_1-{\bf R}_2)=\frac{e^2}{\ell}\frac{N_{m_1m_2}({\bf R}_1-{\bf R}_2)}{D_{m_1m_2}({\bf R}_1-{\bf R}_2)}
\label{Uwvfct}
\end{equation}
where the functions $N_{m_1m_2}$ and $D_{m_1m_2}$ are given, for $n=2$ and $m_1,m_2=0,1$, by
($\mbox{I}_0$ and $\mbox{I}_1$ are modified Bessel functions)
\begin{widetext}
\begin{subequations}
\begin{align}
&N_{00}(r)=-\frac{\sqrt{\pi}e^{-\frac{r^2}{4}}}{32768}\big(9360-2944r^2+560r^4-32r^6+r^8\big)
\notag\\
&+\frac{\sqrt{\pi}e^{-\frac{r^2}{8}}}{65536}\Big[\big(18720+992r^2+240r^4+12r^6+r^8\big)\mbox{I}_0
(\frac{r^2}{8})-r^2\big(2432+312r^2+16r^4+r^6\big)\mbox{I}_1(\frac{r^2}{8})\Big],\label{N00}
\\
&D_{00}(r)=1-\frac{e^{-\frac{r^2}{4}}}{16384}\big(128-32r^2+r^4\big)^2,
\\
&N_{11}(r)=-\frac{\sqrt{\pi}e^{-\frac{r^2}{4}}}{2097152}\big(520384-400512r^2+150672r^4-24448r^6+
1964r^8-72r^{10}+r^{12}\big)
\notag\\
&+\frac{\sqrt{\pi}e^{-\frac{r^2}{8}}}{4194304}\Big[\big(1040768+27840r^2+47120r^4-6656r^6+904r^8-44r^{10}
+r^{12}\big)\mbox{I}_0(\frac{r^2}{8})
\notag\\
&-r^2\big(169088+35760r^2-3904r^4+752r^6-40r^8+r^{10}\big)\mbox{I}_1(\frac{r^2}{8})\Big],\label{N11}
\\
&D_{11}(r)=1-\frac{e^{-\frac{r^2}{4}}}{1048576}\big(-1024-384r^2-40r^4 + r^6 \big)^2, 
\\
&N_{01}(r)=-\frac{\sqrt{\pi}
e^{-\frac{r^2}{8}}}{262144}\Bigg\{e^{-\frac{r^2}{8}}\big(10784+16144r^2-5152r^4
+816r^6-46r^8+r^{10}\big)\notag\\
&-\frac{1}{2}\Big[(128192+16976r^2+464r^4+268r^6-12r^8+r^{10})\mbox{I}_0\big(\frac{r^2}{8}\big)
\notag\\
&-r^2(24976+1440r^2+244r^4-8r^6+r^8)\mbox{I}_1\big(\frac{r^2}{8}\big)\Big]\Bigg\}, \label{N01}
\\
&D_{01}(r)=1-\frac{e^{-\frac{r^2}{4}}r^2}{262144}{\big(128-32r^2+r^4\big)}^2.
\label{D01}
\end{align}
\end{subequations}
\end{widetext}

A plot of the function $U_{01}(r)$ (see Fig.~\ref{figU01wvfct}) shows the same qualitative behavior as the
one obtained within the Landau gauge, Eq.~(\ref{Ueff}),
with, namely, a negative curvature at the origin and a local minimum
at a finite value $r=r_0'\simeq 1.41\ell$. Because the wavefunctions $\varphi_{0}({\bf r})$ and
$\varphi_{1}({\bf r})$ are orthonormal, the denominator $D_{01}(r)$ is always very close to unity, and the
potential $U_{01}(r)$ is very well approximated by the numerator $N_{01}(r)$. The situation is, however,
markedly different for the pairs of wavefunctions $\varphi_{0}({\bf r}-{\bf R})$ and $\varphi_{0}({\bf r}-{\bf R}')$,
and $\varphi_{1}({\bf r}-{\bf R})$ and $\varphi_{1}({\bf r}-{\bf R}')$, which have very strong overlap at
small separations $|{\bf R}-{\bf R}'|$, with the consequence that the denominators $D_{00}(r)$ and
$D_{11}(r)$ become very small for small $r$, leading to nonsensical results for the interaction potentials
$U_{00}(r)$ and $U_{11}(r)$ (see Fig.~\ref{figU00wvfct}). As it can be seen in Figs.~(\ref{figU00wvfct},
 \ref{figU11wvfct}), we find that a good approximation to the true Hartree-Fock interaction potentials
$U_{00}(r)$ and $U_{11}(r)$ is obtained by keeping the numerator contributions only. 
Similar results are obtained for the interaction potentials $U_{02}(r)$, $U_{12}(r)$, $U_{22}(r)$, but
we choose not to write down the explicit expressions of these potentials here for brevity.

\begin{figure}[t]
\centering
\includegraphics[width=8.09cm, height=5cm]{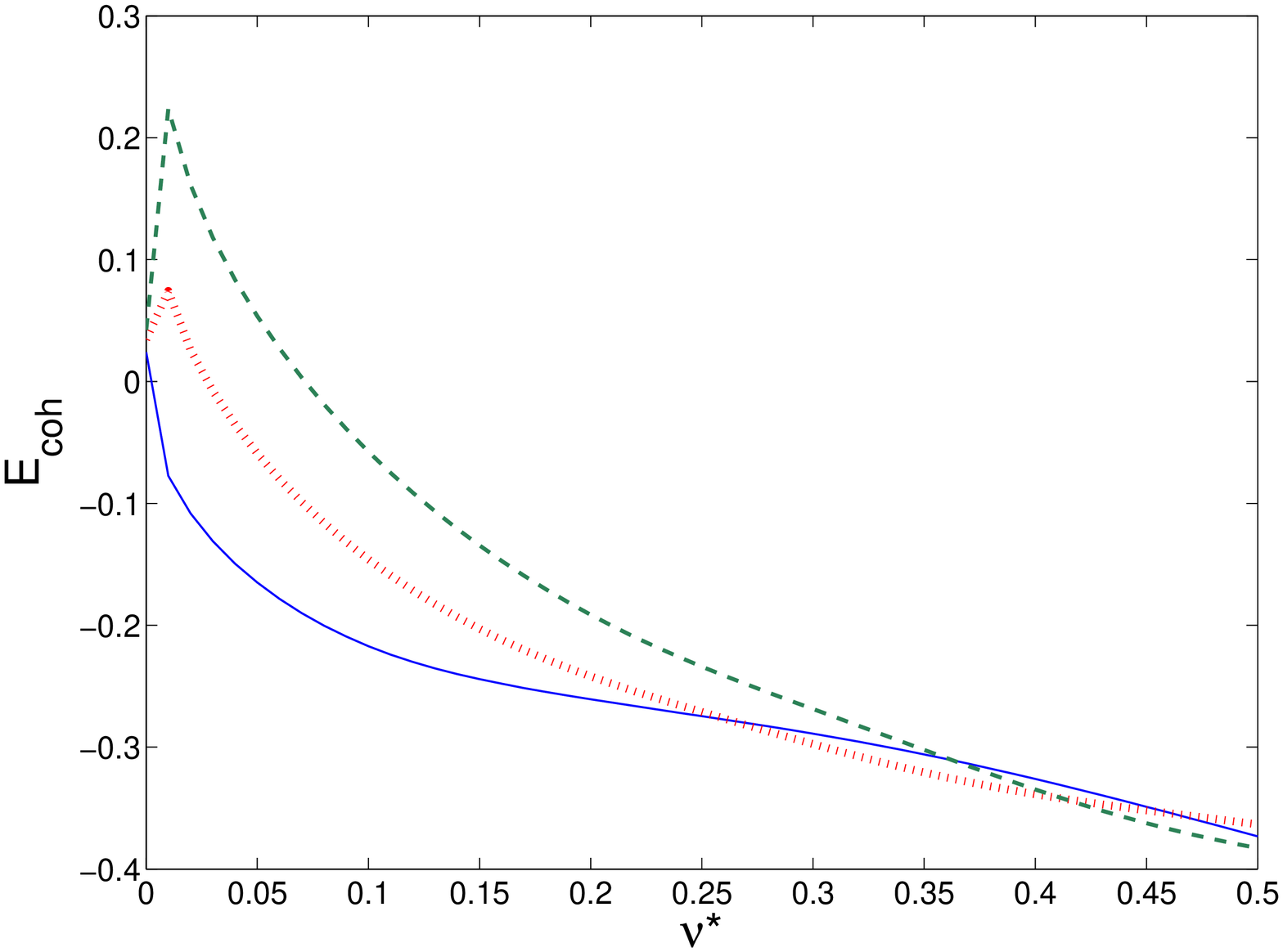}
\caption{(Color online) Cohesive energy (in units of $e^2/\ell$) vs. $\nu^*$ for the
Wigner crystal (solid line), for the 2e bubble solid (dotted line), 
and for the 3 electron bubble crystal (dashed line) in the $n=2$
Landau level, as obtained from the wavefunction approach of this Appendix,
which is based on the symmetric gauge. The relative behavior of these cohesive energies
is very similar to the behavior derived in the approach of the text based on the Landau gauge.}
\label{Fig:PhaseDiagSymmGauge}
\end{figure}

Using the above effective interaction potentials between guiding centers leads to the cohesive energies 
for the Wigner and $M=1$ and 2 bubble crystals in LL $n=2$ shown
in Fig. \ref{Fig:PhaseDiagSymmGauge}. As it can be seen,
the qualitative picture of the relative behavior of these cohesive energies
is the same as in the approach of the text, which is based on the Landau gauge.
We thus see that the phase diagram of electrons in higher LLs is quite robust,
and that a particular choice of gauge is likely not to have an effect on the
qualitative results of HF calculations, even though slight
quantitative discrepancies between calculations carried out in different gauges
might exist.

\section{Numerical values of physical quantities}
\label{appNumerics}

In this Appendix, we give a summary of numerical values of physical quantities we use to calculate the normal
modes in Sec. \ref{NormalModes}. Throughout this Appendix, we shall use CGS units, where the speed of light
in vacuum $c$, Planck's constant $h$, the electron charge $e$ and effective mass in the host semiconductor
$m^*$ are given by
\begin{subequations}
\begin{eqnarray}
c & = & 3\times 10^{10} \;\mbox{cm/s},\\
h & = & 6.6262\times 10^{-27} \;\mbox{erg}\cdot \mbox{s},\\
e & = & 4.80\times 10^{-10} \;\mbox{esu},\\
m^* & = & 0.067 \times 9.11\times 10^{-28}\;\mbox{g} 
\nonumber\\
&=& 6.103\times 10^{-29}\;\mbox{g}.
\end{eqnarray}
\end{subequations}
Following Ref.~\onlinecite{Cote2003}, for the dielectric constant of the host semiconductor, we use the value
$\epsilon= 12.9$, relevant to GaAs, and take for the electron density the typical value
$n_0=3.2\times 10^{11} e^{-}/cm^2$. The resulting expression of the magnetic field $B=n_0hc/e\nu$ in terms
of the filling factor $\nu$ is as follows
\begin{equation}
B = \frac{13.24}{\nu} \times 10^4\mbox{Gauss}.
\label{Bfield}
\end{equation}
The magnetic length $\ell=\sqrt{\hbar c/eB}$ is then
\begin{equation}
\ell = 0.7052\times 10^{-6}\sqrt{\nu}\;\mbox{cm},
\end{equation}
while the effective Bohr radius $a_B=\hbar^2\epsilon/m^*e^2$ is given by
\begin{equation}
a_B = 1.0188\times 10^{-6}\mbox{cm},
\end{equation}
so that the ratio $a_B/\ell$ is given by
\begin{equation}
\frac{a_B}{\ell} = \frac{1.4426}{\sqrt\nu}.
\end{equation}
Now, the electrostatic energy scale $e^2/\epsilon\ell$ is such
that\cite{typo}
\begin{equation}
\frac{e^2}{\hbar\epsilon\ell} = \frac{24046.7}{\sqrt{\nu}}\;\mbox{GHz}.
\end{equation}
Using Eq.~(\ref{Bfield})
we obtain that the cyclotron frequency $\omega_c=eB/m^*c$ is given by
\begin{equation}
\omega_c = \frac{34741.2}{\nu}\;\mbox{GHz},
\end{equation}
so that the ratio of the electrostatic to magnetic energy scales is given by
\begin{equation}
\frac{(e^2/\epsilon\ell)}{\hbar\omega_c} = 0.692\sqrt{\nu}.
\end{equation}

\end{document}